\documentclass[journal=jacsat,manuscript=article,layout=twocolumn]{achemso}
\usepackage{chemformula} % Formula subscripts using \ch{}
\usepackage[T1]{fontenc} % Use modern font encodings
\usepackage{amssymb}     % allow writing "gtrsim"
\usepackage[colorlinks=true, allcolors=blue]{hyperref} % links in references
\usepackage{multirow}    % used in table

\usepackage{xcolor}
\usepackage{graphicx}
\usepackage{float}
\usepackage{xspace}

\usepackage{siunitx}
\DeclareSIUnit[number-unit-product=]\percent{\char`\%} % remove space before percentage "units"

\let\Im\relax\DeclareMathOperator{\Im}{Im}
\usepackage[inline]{enumitem}

\def\ii{{\rm i}}        \def\ee{{\rm e}}
\def\Ree{{\rm Re}}      
\def\rb{{\bf r}}        \def\Rb{{\bf R}}

          \def\kparb{{\bf k}_\parallel}
\def\Qb{{\bf Q}}                
\def\Hh{\mathcal{H}}
\def\me{m_{\rm e}}      
          \def\EF{{E_{\rm F}}}
\newcommand{\ep}{\epsilon}
\def\eps{\epsilon}    % definition of the permittivities
\def\en{\varepsilon}  % definition of the electron frequencies
\newcommand{\pll}{\parallel}
\def\dper{d_{\perp}}  
\def\dpar{d_{\parallel}}
\def\nm{{\rm nm$^{-1}$}}
\def\rdm{r_{\rm dm}}    \def\rmd{r_{\rm md}}
\def\tdm{t_{\rm dm}}    \def\tmd{t_{\rm md}}
\def\rgr{r_{\rm g }}    \def\tgr{t_{\rm g }}
      
\def\epsm{\eps_{\rm m}} \def\epsd{\eps_{\rm d}}
\def\epsb{\eps_{\rm b}}
       \def\wp{\omega_{\rm p}}
      
\def\phiind{\phi^{\rm ind}}    \def\phiext{\phi^{\rm ext}}

\author{A.~Rodr\'{\i}guez~Echarri}
\affiliation{ICFO --- Institut de Ci{\`e}ncies Fot{\`o}niques, The Barcelona Institute of Science and Technology, 08860 Castelldefels (Barcelona), Spain}
\alsoaffiliation{Center for Nano Optics, University of Southern Denmark, Campusvej~55,~DK-5230~Odense~M, Denmark}
\author{P.~A.~D.~Gon\c{c}alves}
\affiliation{Center for Nano Optics, University of Southern Denmark, Campusvej~55,~DK-5230~Odense~M, Denmark}
\author{C.~Tserkezis}
\affiliation{Center for Nano Optics, University of Southern Denmark, Campusvej~55,~DK-5230~Odense~M, Denmark}
\author{F.~Javier~Garc\'{\i}a~de~Abajo}
\affiliation{ICFO --- Institut de Ci{\`e}ncies Fot{\`o}niques, The Barcelona Institute of Science and Technology, 08860 Castelldefels (Barcelona), Spain}
\alsoaffiliation{ICREA --- Instituci\'o Catalana de Recerca i Estudis Avan\c{c}ats, Passeig~Llu\'{\i}s~Companys~23, 08010 Barcelona, Spain}
\author{N.~Asger~Mortensen}
\affiliation{Center for Nano Optics, University of Southern Denmark, Campusvej~55,~DK-5230~Odense~M, Denmark}
\alsoaffiliation{Danish Institute for Advanced Study, University of Southern Denmark, Campusvej~55,~DK-5230~Odense~M, Denmark}
\author{Joel~D.~Cox}
\affiliation{Center for Nano Optics, University of Southern Denmark, Campusvej~55,~DK-5230~Odense~M, Denmark}
\alsoaffiliation{Danish Institute for Advanced Study, University of Southern Denmark, Campusvej~55,~DK-5230~Odense~M, Denmark}
\email{cox@mci.sdu.dk}

\title{Optical response of noble metal nanostructures: Quantum surface effects in crystallographic facets}

\abbreviations{IR,NMR,UV}

\begin{document}

% \begin{tocentry}
% 
% Some journals require a graphical entry for the Table of Contents.
% This should be laid out ``print ready'' so that the sizing of the
% Inside the \texttt{tocentry} environment, the font used is Helvetica
% 8\,pt, as required by \emph{Journal of the American Chemical
% The surrounding frame is 9\,cm by 3.5\,cm, which is the maximum
% permitted for  \emph{Journal of the American Chemical Society}
% graphical table of content entries. The box will not resize if the
% content is too big: instead it will overflow the edge of the box.
% 
% This box and the associated title will always be printed on a
% separate page at the end of the document.
% 
% \end{tocentry}
\begin{abstract}
Noble metal nanostructures are ubiquitous elements in nano-optics, supporting plasmon modes that can focus light down to length scales commensurate with nonlocal effects associated with quantum confinement and spatial dispersion in the underlying electron gas. Nonlocal effects are naturally more prominent for crystalline noble metals, which potentially offer lower intrinsic loss than their amorphous counterparts, and with particular crystal facets giving rise to distinct electronic surface states. Here, we employ a quantum-mechanical model to describe nonclassical effects impacting the optical response of crystalline noble-metal films and demonstrate that these can be well-captured using a set of surface-response functions known as Feibelman $d$-parameters. In particular, we characterize the $d$-parameters associated with the (111) and (100) crystal facets of gold, silver, and copper, emphasizing the importance of surface effects arising due to electron wave function spill-out and the surface-projected band gap emerging from atomic-layer corrugation. We then show that the extracted $d$-parameters can be straightforwardly applied to describe the optical response of various nanoscale metal morphologies of interest, including metallic ultra-thin films, graphene-metal heterostructures hosting extremely confined acoustic graphene plasmons, and crystallographic faceted metallic nanoparticles supporting localized surface plasmons. The tabulated $d$-parameters reported here can circumvent computationally expensive first-principles atomistic simulations to describe microscopic nonlocal effects in the optical response of mesoscopic crystalline metal surfaces, which are becoming widely available with increasing control over morphology down to atomic length scales for state-of-the-art experiments in nano-optics.
\end{abstract}

\section{Introduction}

Metals support collective oscillations of their conduction electrons, known as plasmons, with light-trapping and manipulation capabilities at nanometer length scales, i.e., well below the diffraction limit imposed by traditional optics~\cite{Gramotnev:2010,Gramotnev:2014}. The wealth of fundamental explorations in plasmonics over the last couple of decades has contributed to shape the field of nano-optics~\cite{Fernandez-Dominguez:2017}, holding great promises for light-based technologies including theranostics~\cite{Lal:2008,Rastinehad:2019}, photocatalysis~\cite{WHD12,Zhang:2013}, plasmonic colors~\cite{Kristensen:2017}, solar energy harvesting~\cite{Atwater:2010,SFJ15}, and quantum information~\cite{BA11,paper327,Fernandez-Dominguez:2018}. Advances in modern nanofabrication techniques have driven plasmonics research further by enabling the realization of nano-optical devices that operate on deep-subwavelength scales~\cite{Zhou:2016,Baumberg:2019}. As current capabilities can routinely pattern metallic nanostructures down to the few-nanometer regime~\cite{KMS14,paper326,paper335}, where the frontiers of quantum and classical physics coalesce, new routes towards next-generation plasmon-based technologies begin to emerge, while also posing new challenges in understanding and modeling their optical response at nanometric scales~\cite{BS19,Fernandez-Dominguez:2018}.

The realization of thin crystalline noble-metal films~\cite{Huang:2010,Hoffmann:2016,Mejard:2017,Cheng:2019} is key to cutting-edge explorations of novel plasmonic devices: metallic nanostructures with a high degree of crystallinity are anticipated to suffer reduced Ohmic losses when compared to their amorphous counterparts~\cite{MJK15}, with the recent observation of plasmons in few-atom-thick crystalline silver films partially confirming this intuitive concept~\cite{paper335}. Furthermore, it is well-established in surface science~\cite{Inglesfield:1982} that (111) noble metal surfaces possess Shockley surface states~\cite{Shockley:1939}, whose features resemble those of a two-dimensional electron gas (2DEG). Shockley surface states can support 2D-like plasmon modes~\cite{Echenique:2001} that can be characterized, e.g., by angle-resolved spectroscopy~\cite{Suto:1989,Diaconescu:2007,Politano:2015}, while they can potentially play a role in near-field light--matter interactions at such surfaces.

First-principles simulation methods capture nonclassical effects in the optical response of ultra-thin metal films or few-atom metal clusters~\cite{Varas:2016,Zhu:2016}, but necessitate intensive computational effort that rapidly becomes unfeasible for structures with characteristic sizes $\gtrsim \SI{10}{\nm}$; unfortunately, precision within $\sim10$ is what is currently afforded by state-of-the-art top-down nanofabrication techniques. One of the overarching challenges in theoretical nano-optics is thus to describe the optical properties of nanostructured metals by solving Maxwell's equations while accounting---in the response functions entering the constitutive relations---for quantum-mechanical effects that emerge when electrons are confined in low-dimensional systems, ideally without resorting to overly-demanding numerical approaches that obscure the underlying physics. Fortunately, the situation is somewhat simplified in metals by their ability to effectively screen electromagnetic fields, which leads to an optical response dominated by surface effects. In this context, the concept of microscopic surface-response functions, like the Feibelman $d$-parameters~\cite{F1982,CYA17,GCR20,Goncalves_SpringerTheses,Liebsch:1997,Goncalves_arxiv}, offers a practical and scalable recipe to simultaneously incorporate quantum mechanical phenomena, such as electronic spill-out, nonlocal effects, and surface-enabled Landau damping, into the optical response of metal nanostructures~\cite{F1982,CYA17,GCR20,Goncalves_SpringerTheses,Yan:2015}, as has been recently demonstrated experimentally~\cite{YZY19}.

Here, we apply a quantum-mechanical framework that describes the metal as a vertical stack of (homogeneous) atomic layers which are modeled as a realistic one-dimensional (1D) potential. The wave functions obtained by solving the corresponding Schr\"{o}dinger equation are then used to compute the nonlocal optical response of selected noble metal (gold, silver, and copper) films with specific crystallographic orientations---namely, the (100) and (111) facets---from which we extract the associated Feibelman $d$-parameters. We demonstrate that the $d$-parameters obtained for a thick film (tantamount to a semi-infinite metal), once tabulated (as we do here), can be straightforwardly incorporated into a wide range of electromagnetic problems, ranging from analytical solutions for simple geometries to full-wave numerical electromagnetic solvers of realistic particles, to accurately describe intrinsic quantum mechanical effects affecting the system's optical response~\cite{GCR20,YZY19}. We anticipate that the results presented herein can be widely deployed to describe ongoing experiments and engineer future nanoscale plasmonic devices.

\section{Results and discussion}

Classically, the metal response to a monochromatic electromagnetic field of frequency $\omega$ can be described by the Drude model through its local-response dielectric function~\cite{GarciadeAbajo:2008}
\begin{equation}\label{eq:Drude}
    \epsm(\omega) = \epsb(\omega) - \frac{\wp^2}{\omega(\omega + \ii\gamma^{\rm exp})} ,
\end{equation}
where $\epsb(\omega)$ is an \textit{ad hoc} correction that accounts for the screening associated with the core electrons in the metal, while the second term describes the response of free electrons characterized by a plasma frequency $\wp$ and a phenomenological scattering rate $\gamma^{\rm exp}$ typically extracted from experiments. To maintain fidelity with experimental data, we follow ref~\citenum{GarciadeAbajo:2008} and construct $\epsb(\omega)$ by subtracting the free-electron component from the experimentally tabulated dielectric function~\cite{JC1972}, $\epsm^{\rm exp}$, so that $\epsb(\omega)=\epsm^{\rm exp}(\omega)+(\wp^{\rm exp})^2/(\omega^2 + \ii \omega \gamma^{\rm exp})$; the parameters used to characterize the noble metals that will be considered in this work, namely silver (Ag), gold (Au), and copper (Cu), are specified in Table~\ref{Drude_table} below.

\begin{table}
\centering
\begin{tabular}{c|cc}  \hline
Material  & $ \hbar \wp^{\rm exp}$ (eV) & $\hbar \gamma^{\rm exp}$ (eV) \\ \hline
Ag        &      9.17                  & 0.021  \\ 
Au        &      9.06                  & 0.071  \\  
Cu        &      8.88                  & 0.103  \\ \hline      
\end{tabular}
\caption{{\bf Characterization of the free electron gas.} Parameters used in the Drude dielectric function obtained from experimental data \cite{JC1972}. }
\label{Drude_table}
\end{table}

In classical electrodynamics, a metal surface is commonly described by a dielectric function that changes abruptly from the bulk, local dielectric function of the metal, $\epsm \equiv \epsm(\omega)$, to that of the adjacent dielectric, $\epsd \equiv \epsd(\omega)$. However, this naive prescription can be augmented with $d$-parameter-corrected boundary conditions that incorporate quantum effects associated with the optical response of metal surfaces. Specifically, for a $p$-polarized electromagnetic field impinging on a metal surface from the dielectric side, the nonretarded reflection and transmission coefficients read~\cite{CYA17,GCR20,Goncalves_SpringerTheses,Liebsch:1997} 
\begin{subequations}
\label{eq:rt12}
\begin{align}
    \rdm =& \frac{\epsm-\epsd+(\epsm-\epsd)Q(\dper+\dpar)}{\epsm+\epsd-(\epsm-\epsd)Q(\dper-\dpar)}, \\[0.25em]
    \tdm =& \frac{2\epsd}{\epsm+\epsd-(\epsm-\epsd)Q(\dper-\dpar)} ,
\end{align}
\end{subequations}
respectively. Here, $Q$ is the in-plane wave vector (i.e., parallel to the interface), while $\dper$ and $\dpar$ denote the frequency-dependent, complex-valued quantum surface-response functions introduced by Feibelman~\cite{F1982,CYA17,GCR20,Goncalves_SpringerTheses} (see Methods). On the other hand, for $p$-polarized light impinging on the interface from the metal side, the corresponding nonretarded reflection and transmission coefficients read 
\begin{subequations}
\label{eq:rt21}
\begin{align}
    \rmd =& \frac{\epsd-\epsm+(\epsm-\epsd)Q(\dper+\dpar)}{\epsm+\epsd-(\epsm-\epsd)Q(\dper-\dpar)}, \\
    \tmd =& \frac{2\epsm}{\epsm+\epsd-(\epsm-\epsd)Q(\dper-\dpar)} .
\end{align}
\end{subequations}
Note that, in general, $\rmd \neq -\rdm$ when $d_{\alpha}\neq 0$, where $\alpha=\perp,\pll$.
Equipped with eqs~\ref{eq:rt12} and \ref{eq:rt21}, the overall Fabry--P\'erot (FP)-like reflection coefficient of the composite dielectric--metal--dielectric heterostructure can be determined via
\begin{equation}
    R = \rdm + 
    \frac{\tdm \, \tmd \, \rmd \, \ee^{-2 Q L}}{1 - \rmd \, \rdm \, \ee^{-2 Q L}},
    \label{eq:FP_equation}
\end{equation}
where $L$ denotes the metal film thickness.

\begin{figure*}[t]
\centering
\includegraphics[width=1.0\textwidth]{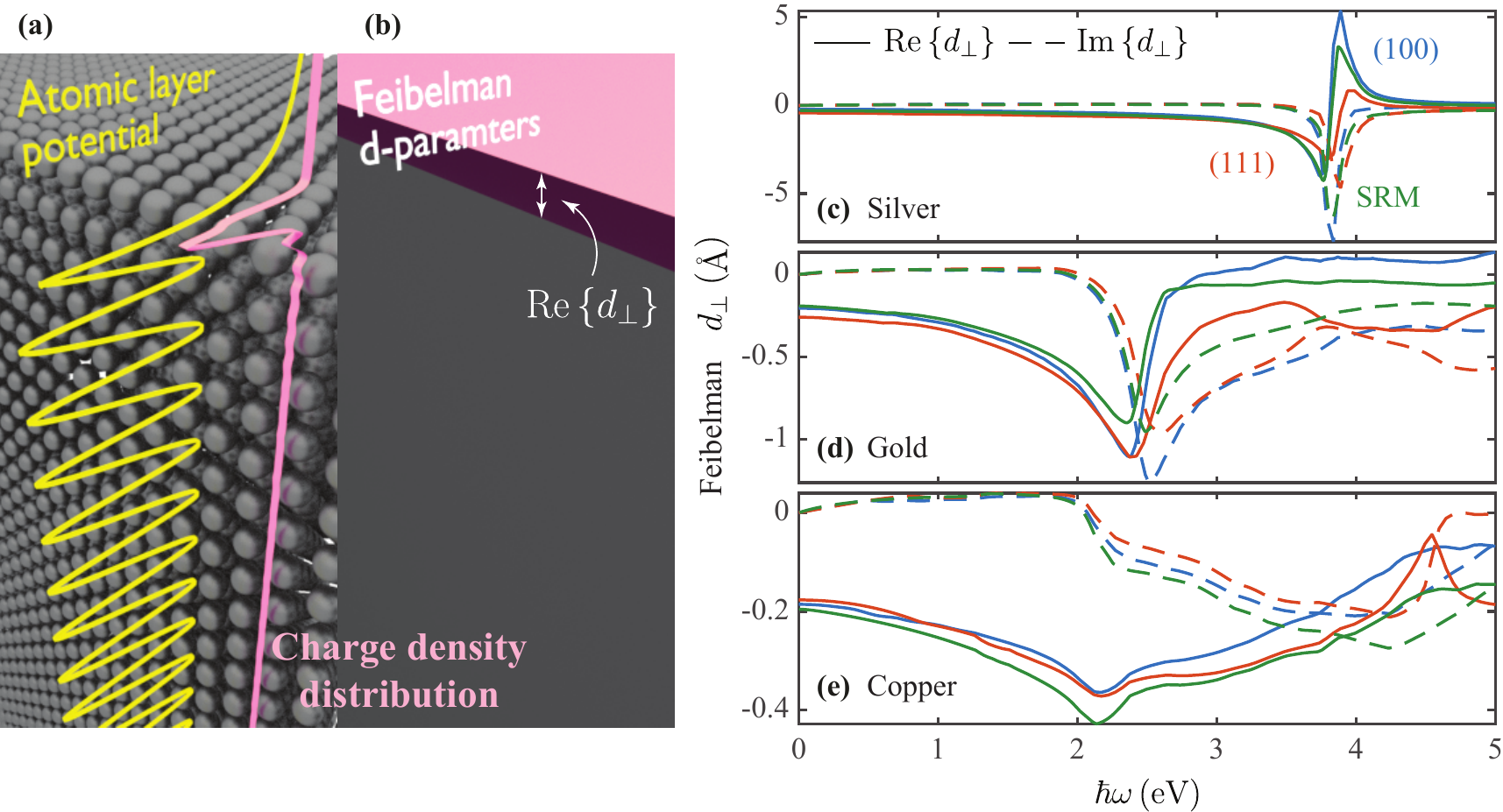}
\caption{{\bf Feibelman \emph{d}-parameters for noble metals.} We illustrate in panel (a) a semi-infinite crystalline noble metal surface comprised of vertically-stacked atomic planes and characterized by a phenomenological atomic layer potential (ALP), for which the optical response is computed in atomistic quantum-mechanical simulations; extraction of the Feibelman \emph{d}-parameters associated with the metal surface facilitates a surface-corrected classical treatment of its optical response, portrayed as the plane in panel (b) representing the centroid of the induced charge, that incorporates quantum nonlocal effects. In panels (c-e) we present the real (solid curves) and imaginary (dashed curves) parts of extracted $\dper$ for the (111) and (100) facets of Ag, Au, and Cu, along with $\dper$ obtained in the SRM, as indicated by the color-matched legends.}
\label{fig_feibelman}
\end{figure*}

The above quantum-surface-corrected description only requires as inputs the Feibelman response functions, $\dper$ and $\dpar$, which can be obtained for a particular dielectric--metal interface either experimentally~\cite{YZY19} or from sophisticated quantum mechanical methods [e.g., time-dependent density-functional theory (TDDFT) or empirical quantum-corrected models]. Here we obtain the \emph{d}-parameters associated with crystalline noble metal surfaces from nonclassical optical response simulations of their reflection coefficients based on the random-phase approximation (RPA), employing the QM model reported in ref~\citenum{paper329} and detailed in the Methods section. More specifically, we construct the RPA susceptibility from single-particle wave functions $\Psi_{i}(\rb)$ satisfying the Schr\"{o}dinger equation for a one-dimensional (1D) phenomenological potential $V(z)$ characterizing each material~\cite{CSE99}; this so-called atomic layer potential (ALP) is fitted to reproduce salient features of the bulk and semi-infinite surface electronic structures, so that the ALP-RPA description captures the effects of electronic band structure, electron spill-out, binding energies, and transverse atomic corrugation in the optical response of layered Ag, Au, and Cu films with (100) or (111) crystallographic orientation (see Table~\ref{bands_paramters}). See the Methods section for a comprehensive description of the calculation, in which the metal background polarizability $\epsb$ incorporating core electron screening is also included when describing the electron-electron interaction.

Using the ALP-RPA model, we follow the prescription outlined in the Methods section to extract the Feibelman $d$-parameters presented in Figure~\ref{fig_feibelman}. The $\dper$ associated with crystalline noble metal surfaces are contrasted with those obtained within the specular reflection model (SRM), first proposed by Ritchie and Marusak to study nonlocal effects on surface plasmon dispersion~\cite{RM1966}, and later generalized by others to deal with more complex structures~\cite{RM1966,Ford:1984,Pitarke:2007,GarciadeAbajo:2008,Goncalves_SpringerTheses}; the SRM---also known as the semiclassical infinite barrier model (SCIB)---incorporates bulk spatial dispersion (i.e., nonlocality) but assumes a homogeneous electron gas and thus neglects atomic corrugations. In the case of the (100) orientation and the SRM, the absence of surface currents for charge-neutral materials fixes $\dpar=0$~\cite{F1982}, while $\dpar$ is introduced heuristically in order to incorporate the response of the Shockley surface states in the (111) facet; as explained in Methods, when extracting $\dper$ for the (111) facets we explicitly omit intraband transitions involving surface states to avoid double-counting the effect of the 2DEG. The Feibelman $\dper$-parameters presented in Figure~\ref{fig_feibelman} clearly distinguish the surface response functions of a metal's crystallographic facets. In particular, there are two main regions at low energies ($\hbar \omega<1$ eV) and at energies around $ \hbar \omega_{\text{sp}}^{\text{cl}} = \hbar\omega_{\text{p}}/\sqrt{\epsb(\omega_{\text{sp}}^{\text{cl}}) + \epsd}$ where the classical nonretarded surface plasmon is spectrally centered.

% ------------- FIGURE SP dispersion ---------------------
\begin{figure*}[t]
\centering
\includegraphics[width=1.0\textwidth]{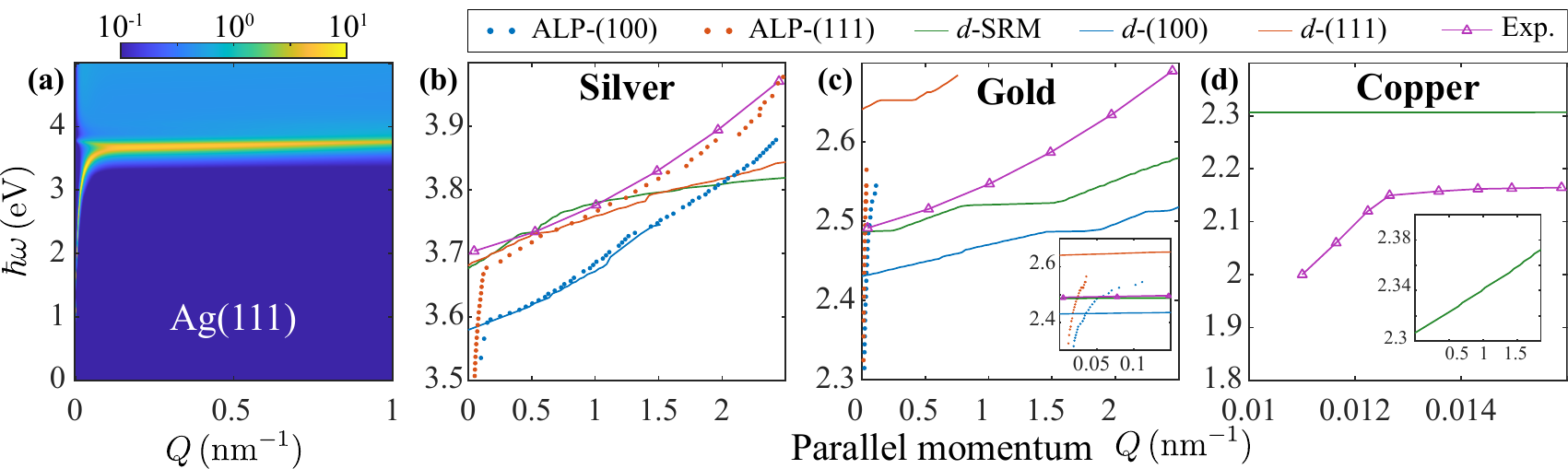}
\caption{{\bf Surface plasmon dispersion.} (a) Loss function, $\Im\{r\}$, of a thick film consisting of $N=100$ Ag(111) layers (i.e., in the semi-infinite limit) simulated with the ALP model. In panels (b-d) we present the dispersion relations of the indicated metals and facets as determined from maxima in their computed loss functions; we extract the maxima directly from $\Im\{r\}$ for ALP calculations (color-coded points), while the dispersion of the $d$-parameter corrected model is given by the poles of the denominator of $r_{\rm dm}$ in eq~\ref{eq:rt12} (solid curves). Triangular symbols represent experimental data (for Ag~\cite{RYB95}, Au~\cite{PP09}, and Cu~\cite{RB1981}).}
\label{fig_SP} 
\end{figure*}
% ------------- FIGURE SP dispersion ---------------------

\hfill

\textbf{Nonretarded surface plasmon dispersion.} %
Equipped with the $d$-parameters of various noble metals and their crystal facets, along with the analytical expression of eq~\ref{eq:rt12}, we illustrate their ability to reproduce the nonretarded surface plasmon dispersion (given by the poles of $\rdm$) obtained directly from the ALP. While the ALP method actually describes a crystalline metal film of finite thickness, we obtain well-converged results for $N \gtrsim 50$ monolayers,. Figure~\ref{fig_SP}a shows the loss function $\Im\{r\}$ for silver with a (111) crystallographic orientation in the ALP model as a function of optical in-plane wave vector and energy; peaks in $\Im\{r\}$ indicate the surface plasmon dispersion, which tends toward zero frequency for small $Q$, in agreement with eq~\ref{eq:FP_equation} for a film of finite thickness and non-vanishing $\ee^{-2QL}$. Figures~\ref{fig_SP}b--d display the nonretarded surface plasmon dispersion for Ag, Au, and Cu, obtained from the ALP (colored dots) along with the Feibelman $d_\perp$-parameter (solid curves), together with available experimental data \cite{RB1981,RYB95,PP09}.

Here, the $d_\perp$-parameter-results have been extracted by comparing eq~\ref{eq:rt12} with the ALP reflection coefficient in the thick-film limit; this procedure, however, needs to be carried out judiciously as the conditions $Q L \gg 1$ and $ Q |d_\perp| \ll 1$ must be simultaneously fulfilled. Chiefly, our results show that the optical response of a metal surface is determined by the surface's specific crystallographic orientation and that its surface response can be well-described in terms of the Feibelman $d_\perp$-parameter, as evidenced by the overlapping dispersion relation calculated via the ALP model.

From eq~\ref{eq:rt12}, it follows that the nonclassical surface plasmon dispersion in the nonretarded regime---keeping only terms up to first-order in $q d_\alpha$---exhibits an approximately linear behavior with in-plane wave vector $Q$, namely~\cite{Liebsch:1997,GCR20,Goncalves_SpringerTheses}
\begin{equation}
 \Ree \left\{\omega_{\text{sp}}\right\} \approx \omega_{\text{sp}}^{\text{cl}} \left[ 1 - \frac{\epsd \Ree \left\{d_\perp - d_\pll\right\}}{\epsb(\omega_{\text{sp}}^{\text{cl}}) + \epsd} Q \right], 
\end{equation}
where $\omega_{\text{sp}}^{\text{cl}} \equiv \Ree \big\{ \omega_{\text{p}}/\sqrt{\epsb(\omega_{\text{sp}}^{\text{cl}}) + \epsd} \big\}$ is the classical nonretarded surface plasmon frequency~\cite{Pitarke:2007}.

While the surface-corrected Ag(111) response obtained from both ALP and SRM models is in good agreement with experiment, determination of the gold surface plasmon dispersion is complicated by an immediate onset of broadening in the loss function at low $Q$; the situation is further compounded for copper, where no clear maximum emerges in neither the response described by eq~\ref{eq:rt12} with \emph{d}-parameters nor in the direct ALP calculation, and only the SRM exhibiting well-defined maxima.

\hfill

\begin{figure}[t!]
\centering
\includegraphics{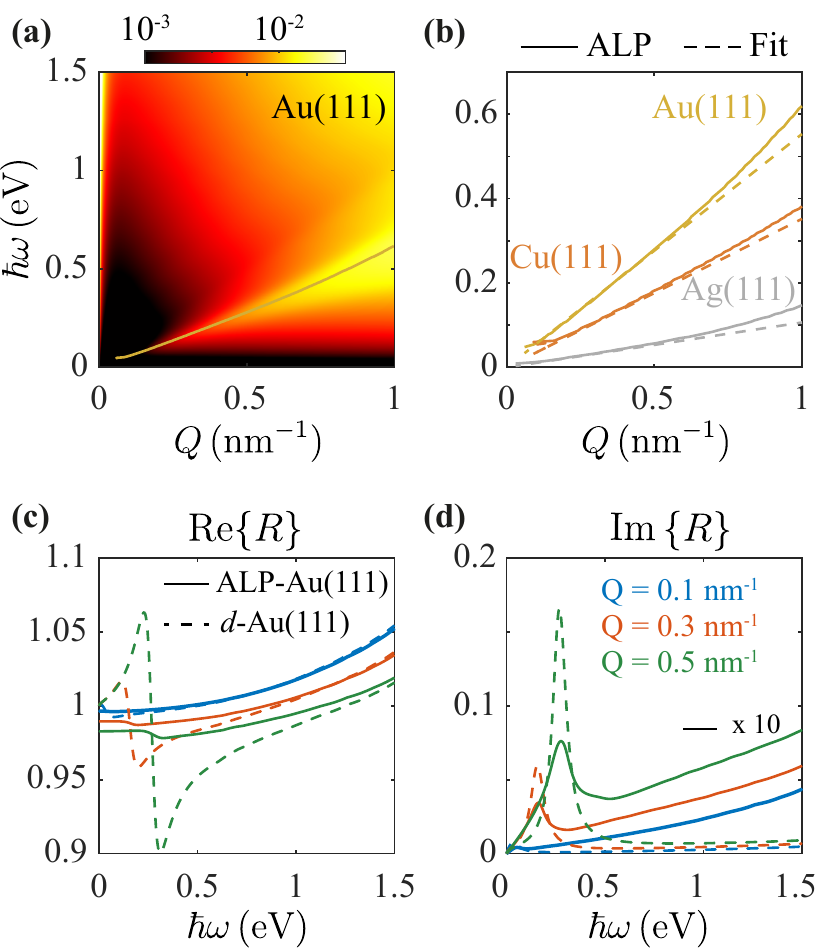}
\caption{{\bf Acoustic surface plasmons on (111) noble metal surfaces.} 
(a) Loss function, $\Im\{r\}$, of a Au(111) surface computed within the ALP model for a thick film in the semi-infinite limit (specifically, $N=100$ layers). 
(b) Acoustic surface plasmon dispersion obtained in the ALP model (solid line) and fitting (dashed line) to the linear dispersion $\omega= v_\phi Q$, with phase velocities $v_\phi$ provided in Table~\ref{bands_paramters}. 
Real (c) and imaginary (d) parts of the reflection coefficient of a Au(111) thick film ($N=100$ layers; indistinguishable from a semi-infinite metal) determined using both the ALP model (solid curves) and the $d$-parameters (dashed curves) for selected parallel wave vectors $Q$ [indicated by the color-matched legend in (d)].}
\label{fig_AP} 
\end{figure}

\textbf{Acoustic surface plasmons due to Shockley surface states.} %
At low frequencies, a feature exhibiting a nearly linear dispersion emerges in the loss function associated with (111)-faceted metal surfaces, indicating the existence of acoustic surface plasmons formed by Shockley surface states~\cite{Pitarke:2004,SPC05,Diaconescu:2007,Pitarke:2007,Pohl:2010,Yan:2012}. Figure~\ref{fig_AP} shows the loss function of a Au(111) surface, which in the low-frequency regime is marked by the presence of a well-defined but relatively broad feature associated with its acoustic surface plasmon (c.f. scales of Figure~\ref{fig_AP}a and Figure~\ref{fig_SP}a). Next, we present the dispersion relation of acoustic surface plasmons akin to the Au(111), Ag(111), and Cu(111) surfaces obtained within the ALP framework (solid curves), and whose slope---corresponding to the acoustic surface plasmon velocity---is then determined through a linear fit (dashed curves); see Table~\ref{SS_table}). Note that the intrinsic acoustic surface plasmons supported by noble metal surfaces of well-defined crystallographic orientation have been characterized experimentally under different conditions: For Au(111), a phase velocity $v_\phi/v_{\rm 2D}=1.7$ was observed at room temperature~\cite{PP10}, while a value $v_\phi/v_{\rm 2D} \approx 0.8$ was reported at $78$\ K~\cite{VSL13}; the extracted value in the present work (see Table~\ref{SS_table}) is close to unity.

\begin{table*}
\centering
\begin{tabular}{c|ccccc}  \hline
Material& $ \EF-\en^\perp_{\rm SS}$ (eV) & $m^*$(SS)$/m_e$    & $v_{\rm 2D}/c$         & $v_\phi$/$v_{\rm 2D}$  &   $\hbar \gamma_{\rm 2D}$ (meV)     \\ \hline
Ag(111) &      0.026               & 0.40~\cite{RNS01}  &  $5.04 \times 10^{-4}$ &          1.0690        &   27.5  \\ \hline
Au(111) &      0.525               & 0.26~\cite{RNS01}  &  $2.81 \times 10^{-3}$ &          0.9971        &   83.7  \\ \hline 
Cu(111) &      0.356               & 0.41~\cite{RNS01}  &  $1.84 \times 10^{-3}$ &          0.9676        &   118.1 \\ \hline      
\end{tabular}
\caption{{\bf Characterization of acoustic plasmons originating from Shockley surface states.} We parametrize the (111) surface state (SS) of a specified noble metal by its energy $\en^\perp_{\rm SS}$ below the Fermi energy $\EF$, effective mass $m^*$, and Fermi velocity $v_{\rm 2D}$ (normalized to the speed of light $c$). The resulting acoustic plasmons are characterized by their phase velocity $v_\phi$ and phenomenological damping $\hbar \gamma_{\rm 2D}$. }
\label{SS_table}
\end{table*}

The Shockley 2DEG supported by the (111)-facets can be accounted for through the Feibelman $\dpar$-parameter (since it can mimic a surface conductivity). We exploit this by introducing, in an \emph{ad hoc} fashion, a heuristic expression for $d_\pll$ as described in the Methods section. We emphasize here that, because we account for the 2DEG heuristically, intraband transitions involving surface states are omitted in the ALP-based $\dper$ calculations. Then, in possession of both $\dper$ and $\dpar$, we use eq~\ref{eq:rt12} to reproduce the optical response calculations performed with the ALP model. As observed in Figure~\ref{fig_SLM_graphene}c-d, where the real and imaginary parts of the reflection coefficient for Au(111) are compared for different values of $Q$, the reconstruction is satisfactory for small $Q$, although the amplitudes of these already weak features are not well-reproduced.

\hfill 

\textbf{Nonclassical optical response of ultrathin metal films.} %
The practical utility of the $d$-parameter framework for mesoscale electromagnetism becomes apparent by recognizing that, once obtained for a specific dielectric--metal interface, they can be readily incorporated in a broad range of optical response calculations, either via $d$-parameter-corrected scattering coefficients~\cite{GCR20} or through $d$-parameter-modified boundary conditions.~\cite{Yan:2015,YZY19}

\begin{figure*}[t!]
\centering
\includegraphics{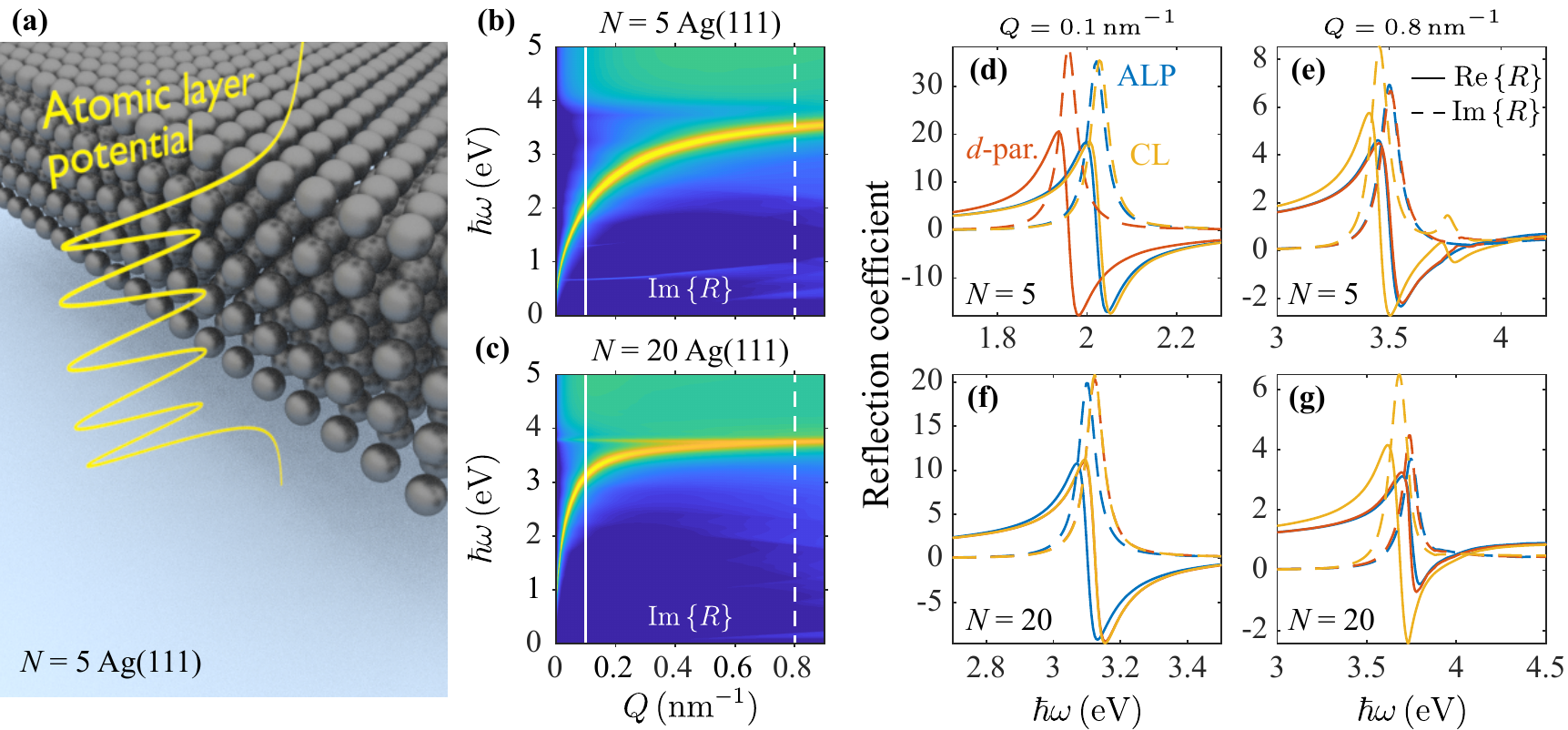}
\caption{{\bf Nonclassical optical response of ultrathin metal films.} (a) graphical illustration of Ag(111) film with $N=5$ monolayers. (b-c) plasmon dispersion relation shown as a feature in the loss function $\Im\{R\}$ for Ag(111) films of (b) $N=5$ and (c) $N=20$ monolayers using the ALP model. White vertical lines in each figure select the loss function at $Q=0.1$ \nm (solid line) and $Q=0.8$ \nm (dashed line), for which (d-e) and (f-g) show the corresponding reflection coefficient $R$ for $N=5$ and $N=20$ monolayers respectively. Colored curves correspond to different models as indicated in the legend in panel (d). In figures (d-g) the reflection coefficient $R$ is computed in the ALP (blue curves) and compared to the Fabry--P\'erot model of Eq.\ \ref{eq:rt12} for cases including the extracted $\dper$ (red curves) and setting $\dper=\dpar=0$ (yellow curves).}
\label{fig_thin_films} 
\end{figure*}

\begin{figure*}[t!]
\centering
\includegraphics[width=1\textwidth]{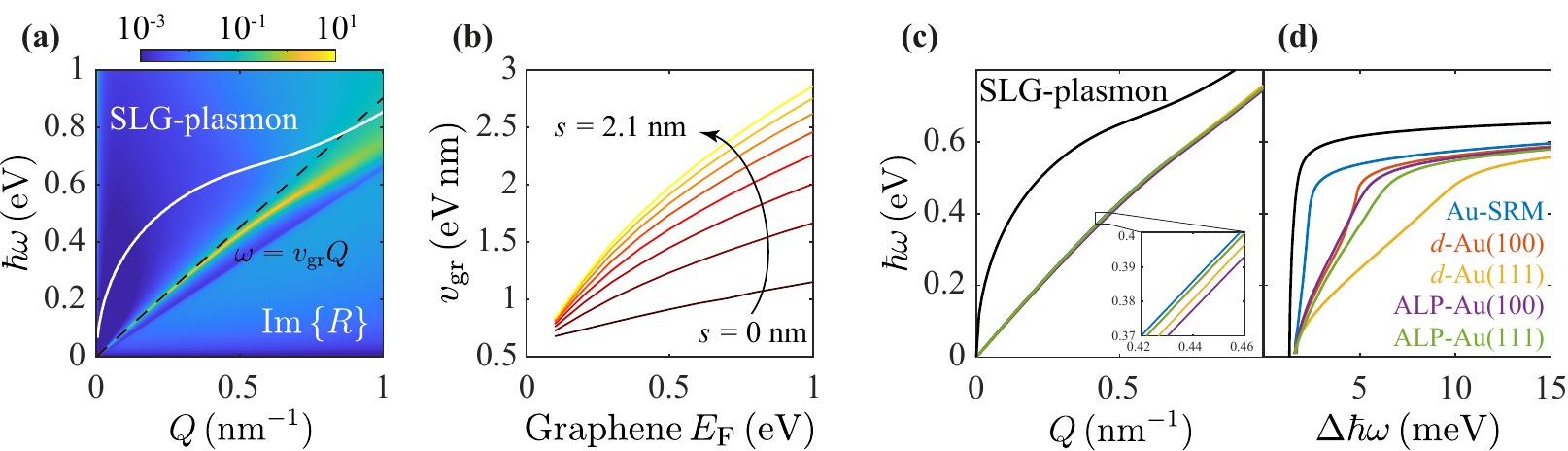}
\caption{{\bf Acoustic graphene plasmons on a Au(111) film.} (a) Loss function $\Im\{R\}$ for single-layer graphene (SLG) doped to a Fermi energy $\EF=0.5$\, eV and deposited directly on top of a semi-infinite metal Au(111) surface, as described with the ALP model. The white curve corresponds to self-standing SLG plasmons, and the black dashed curve is a linear fit to the resulting acoustic plasmon characterized by an acoustic velocity $v_{\rm gr}$. In panel (b) $v_{\rm gr}$ is computed as a function of doping while varying the separation distance from $s=0$ nm to $s=2.1$ nm, in steps of $0.3$ nm. (c) The dispersion relation of the prominent acoustic plasmon feature in (a) is computed using the various gold film models indicated in the legend in (d); the associated linewidths are presented in (d), where the black solid line is a reference for isolated SLG with $\EF=0.5$\, eV.}
\label{fig_SLM_graphene} 
\end{figure*}

As a concrete example, we now investigate the nonclassical optical response of ultrathin silver films comprising $N$ (111) atomic monolayers. In Figure~\ref{fig_thin_films}a-b we present the loss function for ultrathin silver films with thicknesses $N=5$ and $N=20$, which is dominated by the surface plasmon supported by the films; incidentally, the plasmon dispersion relation obtained from the ALP model closely resembles that obtained in a simple Fabry-P\'erot description, even without including $d$-parameters. Figures~\ref{fig_thin_films}(c-f) compare the spectral dependence of the reflection coefficient for selected in-plane wave vectors predicted in the ALP model with FP models that include or neglect the $d$-parameter correction. Here, we observe an unremarkable effect resulting from $\dper$ for small values of the parallel wave vector ($Q=0.1$ \nm), while at larger in-plane momenta, for which the plasmon resonance approaches the plasma frequency ($\omega \to \wp$), noticeable differences emerge (e.g., for $Q=0.8$ \nm). The excellent agreement between the calculated curves based on the $d$-parameter and ALP frameworks underscores how the optical response obtained analytically using eq~\eqref{eq:FP_equation} together with the Feibelman $d$-parameters (c.f.~eqs~\ref{eq:rt12}--\ref{eq:rt21}) accurately accounts for quantum effects impacting the film's electromagnetic response. In particular, the optical response for extremely-thin Ag films, down to $N=5$ atomic planes, appears to be well-reproduced by the surface-corrected thin film reflection coefficient, although the application of the $d$-parameters to describe such films is questionable.

\hfill 

\textbf{Graphene next to crystallographically faceted metal films: Acoustic graphene plasmons} %
We consider the extrinsic acoustic plasmons produced by the hybridization of a closely-spaced graphene layer with a crystalline metal film. Unlike the intrinsic acoustic plasmons supported by the (111)-facets, the introduced graphene layer and its opto-electronic tunability provides an additional knob to actively modulate the optical response of the emerging low-energy acoustic plasmon modes with linearized dispersion~\cite{PVP18}. The experimental capability to position graphene within $\sim1$\,nm of a noble metal layer is launching explorations of extreme light concentration within the gap region~\cite{AND18,LYA19,EAH20}, which could be further improved by employing crystalline noble metals~\cite{paper329}. In what follows, we summarize the semi-analytical FP description of the optical response based on the extracted $d$-parameters.

For a zero-thickness 2D graphene monolayer, the reflection and transmission coefficients in the quasistatic limit~\cite{paper235} read
\begin{align}
r^{\rm 2D}_{\rm g}&=\frac{1}{1-\ii\omega/(2\pi Q\sigma)}, \nonumber\\
t^{\rm 2D}_{\rm g}&=1-r^{\rm 2D}_{\rm g}, \nonumber
\end{align}
where $\sigma(Q,\omega)$ is the nonlocal conductivity of graphene, which we treat here at the level of the nonlocal RPA~\cite{GoncalvesPeres,Goncalves_SpringerTheses,WSS06,HS07} (using Mermin's prescription for the relaxation-time approximation which conserves local particle number~\cite{M1970}; we take $\tau = \SI{500}{\fs}$).

Similarly to eq~\eqref{eq:FP_equation}, we compute the reflection coefficient of an extended graphene sheet on top of a semi-infinite metal via the FP resonance model as
\begin{equation}
    R = \rdm + \frac{ \tgr^2 \, \rdm \, \ee^{-2 Q s}}{1 - \rgr \, \rdm \, \ee^{-2 Q s}},
\end{equation}
where $s$ is the spacing between the metal surface and graphene, and $\rgr$ and $\tgr$ are the reflection and transmission coefficients of graphene. In our calculations we follow the prescription of ref~\citenum{paper329} to account for spatial dependence of the carbon 2p orbitals $\varphi_{\rm 2p}(\rb)$ extending outwards from the graphene monolayer plane, leading to the corrected graphene reflection and transmission coefficients $\rgr=r^{\rm 2D}_{\rm g} C_{\rm g}^2 e^{-Q d_{\rm g}}$ and $\tgr=t^{\rm 2D}_{\rm g} C_{\rm g}^2 e^{-Q d_{\rm g}}$, where $d_{\rm g}=0.33$\,nm is the interlayer spacing of graphite and $C_{\rm g}$ is a coupling factor defined in ref~\citenum{paper329}. Taking into account the aforementioned effective graphene thickness, the separation distance $s$ actually corresponds to the distance between the edge of the graphene and the metal surface, i.e., $s=0$ corresponds to a distance $d_{\rm g}/2$ between the graphene center and the metal surface.

Acoustic plasmons are anticipated to emerge even by depositing graphene directly on metallic films~\cite{PVP18}, heralded by the prominent low-energy linear dispersion feature in the reflection coefficient of Figure~\ref{fig_SLM_graphene}a. For the considered hybrid graphene--Au(111) surface, the excited extrinsic acoustic plasmons are characterized as before by $\omega=v_{\rm gr}Q$ with $v_{\rm gr}$ denoting the associated group velocity; in such a heterostructure, $v_{\rm g}$ is determined by graphene Fermi energy $\EF$ and the graphene--metal spacing $s$, as illustrated in Figure~\ref{fig_SLM_graphene}b. Accordingly, a given dispersion velocity can be obtained by different combinations of $\EF$ and $s$.

Figure~\ref{fig_SLM_graphene}c reveals that among Ag, Au, and Cu (not shown), neither the choice of metal nor the considered crystalline facet strongly influences the acoustic plasmon dispersion characteristics; at low energies, these metals are all good conductors that effectively screen the graphene plasmon and render its dispersion acoustic. We remark, however, that, for the same heterostructure, the dispersion relation of the higher energy plasmon mode is indeed dominated by the metal properties~\cite{paper329}. However, inspection of the linewidths of the acoustic plasmon in Figure.~\ref{fig_SLM_graphene}d reveals a substantial dependence on the quantum-mechanical effects arising from the various crystalline facets, which is underlined by the underestimation of spectral widths in the SRM~\cite{paper329}. In particular, the obtained results suggest that crystalline Au(111) gives rise to additional surface-enhanced damping when compared to Au(100), presumably due to the presence of a surface state, and warranting further study of the acoustic plasmons supported by heterostructured crystalline metal films.

\hfill

\begin{figure}[t!]
\centering
\includegraphics[width=1.0\columnwidth]{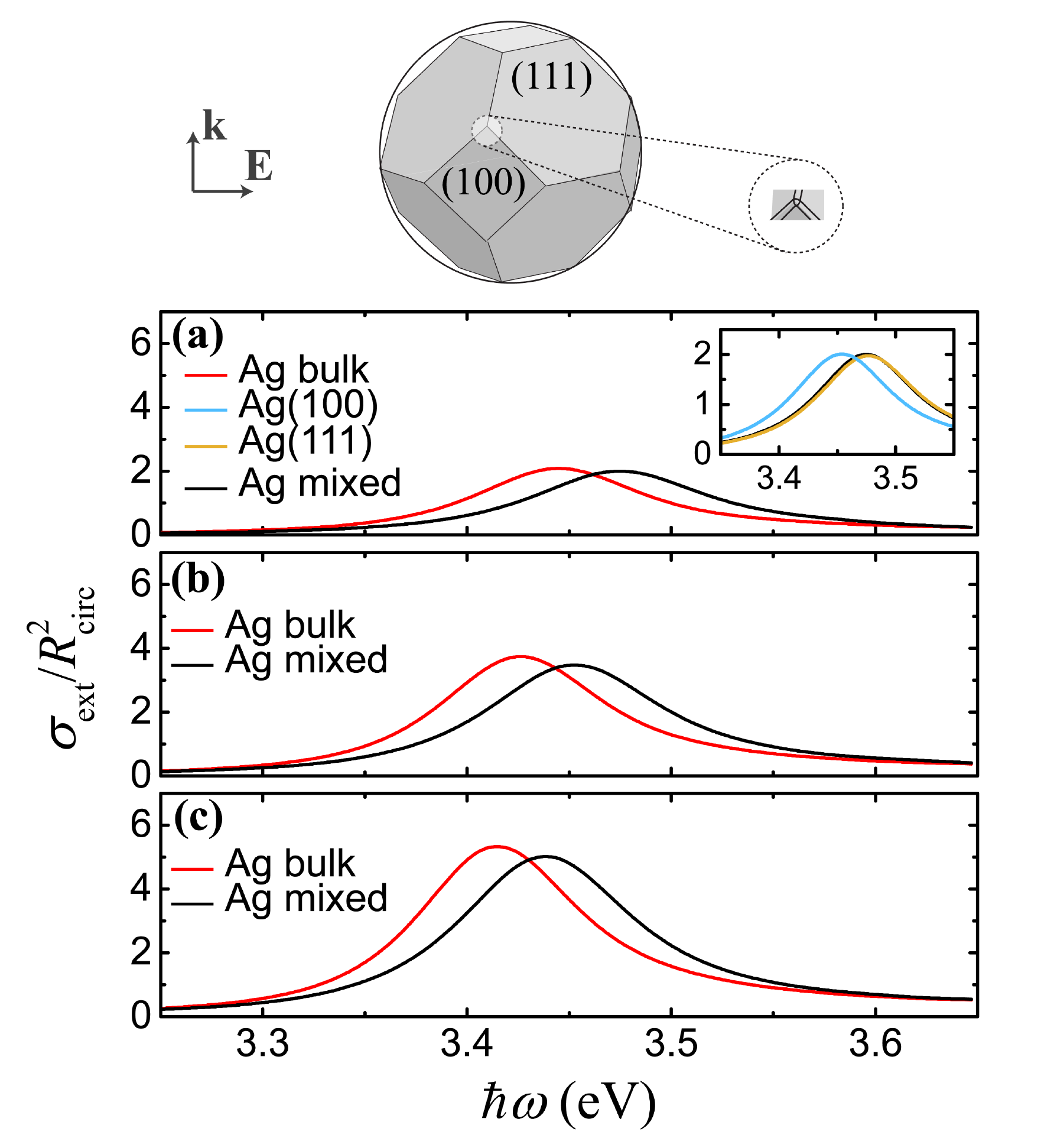}
\caption{{\bf Extinction spectra for a Ag truncated octahedra including quantum surface-corrections.} 
Optical extinction cross section (normalized to the radius of the circumscribed sphere $R_{\mathrm{circ}}$ for Ag truncated octahedra with $R_{\mathrm{circ}} =3$,
$5$, and $7$\;nm (a, b, and c respectively), as shown in the schematics on top. Red curves correspond to the response of bulk Ag (classical with no $d$ parameters),
and black curves to facets $(100)$ and $(111)$ described by their corresponding $d$ parameters shown in Figure~\ref{fig_feibelman}a and b. The blue and yellow curves in the inset of (a) show the corresponding spectra if the entire particle is described entirely by the Feibelman parameters of $(100)$ and $(111)$ facets, respectively.
}
\label{fig_MNP} 
\end{figure}

\textbf{Crystallographically faceted nanoparticles.} %
Going beyond planar, layered media, we explore the role of crystallographic orientation in faceted noble metal nanoparticles (NPs). The optical response of metallic NPs is dominated by the localized surface plasmon (LSP) resonances supported by it; the most prominent of which are typically those of dipolar character, as they can couple to far-field radiation. As the NP size is reduced towards nanometric dimensions, the ensuing NP's surface-to-volume ratio grows and leads to more pronounced nonclassical corrections associated with the NP's quantum surface-response. To illustrate the importance of using the appropriate Feibelman $d$-parameters for determining the quantum surface-response associated with specific facets, we consider in Figure~\ref{fig_MNP} a realistic faceted silver NP. Any natural NP---especially those with characteristic dimensions $\lesssim \SIrange{10}{20}{\nm}$---no matter how carefully synthesized, will always deviate from a perfect sphere, as a consequence of its growth in a sequence of specific crystallographic planes~\cite{Myroshnychenko:2008a}. The shape closest to a sphere is that of a truncated octahedron, as depicted schematically in Figure~\ref{fig_MNP}; it is characterized by large hexagonal $(111)$ surfaces and smaller $(100)$ facets. 

To compute the NP's nonclassical extinction cross-section $\sigma_{\mathrm{ext}}$, we implement in a finite-element method (FEM) solver the mesoscopic boundary conditions within the Feibelman formalism~\cite{Yan:2015,YZY19,GCR20,Goncalves_SpringerTheses}; in practice, this is tantamount to the introduction of \emph{surface} electric and magnetic currents (see Methods) characterized by the $d$-parameters presented in Figure~\ref{fig_feibelman}. Comparing with the classical spectra (red curves), it is clear that the effect of the $d$-parameters is to capture the nonlocal optical response of such an NP. As expected for silver, the spectra are shifted to higher energies as a result of an inward shift ($\Ree\{d_\perp\} < 0$) of the screening charges. Studying different NP sizes, from $3$ to $\SI{7}{\nm}$ in "radius" (meaning here the circumscribed radius $R_{\mathrm{circ}}$), a consistent trend is observed, with the resonance broadening (as a result of increased Landau damping) and undergoing stronger blueshifts, as the size decreases, which is compatible with the predictions of nonlocal hydrodynamics~\cite{GarciadeAbajo:2008,Mortensen:2014,Raza:2015a}, and also in accordance with electron energy-loss spectroscopy (EELS) experiments~\cite{Ouyang:1992,Scholl:2012,Raza:2013,Raza:2015}. The observed behavior is mainly attributed to the $(111)$ facets, as shown in the inset of Figure~\ref{fig_MNP}a, where blue and yellow curves show the corresponding spectra assuming that the entire NP is described solely by the $(100)$ or $(111)$ $d$-parameters, respectively; since the associated surface area of the $(100)$ facets is smaller, and their corresponding Feibelman parameters are significantly lower in magnitude than those of the $(111)$ surface, they only induce a small frequency shift in the spectra. Then, while the main effect is due to the $(111)$ facets, the corresponding spectrum almost coincides with the ``mixed'' one, where each facet is described by its own parameters. It is also worth noting that, because the truncated octahedron constitutes a highly symmetric shape, the optical response of such NPs resembles that of spheres, and thus changing the angle of incidence is not expected to lead to significant differences.

\section{Conclusions}

The inherently large losses sustained by noble metals is often regarded as the Achilles heel of nano-optical functionalities based on plasmonics, and has motivated intensive efforts to identify new material platforms that can support long-lived polaritons. Crystalline noble metal films constitute one appealing possibility that is now becoming increasingly available. We have introduced surface response corrections in the form of Feibelman $d$-parameters extracted from quantum-mechanical optical response calculations of crystalline noble metals. As demonstrated here, the tabulated $d$-parameters for gold, silver, and copper surfaces with specific crystallographic orientations can be straightforwardly incorporated in analytical models as well as in computational electromagnetic solvers for the calculation the nonclassical optical response of various nanoplasmonic systems of current interest that contain these facets. We envision that the $d$-parameters reported here can be widely deployed to describe quantum surface effects in crystalline noble metal surfaces which are actively explored for novel nanophotonic functionalities and applications.

% =================================================
% -------------------- METHODS --------------------
% =================================================
\onecolumn

\section{Methods}

\subsection{Microscopic surface-response functions: Feibelman \emph{d}-parameters}

The surface-response functions introduced by Feibelman~\cite{F1982}, $d_\perp$ and $d_\pll$, are formally given, respectively, by the first-moment of the quantum mechanical induced charge ($\rho_{\text{ind}}$) and of the parallel component of the corresponding current density ($J_x^{\text{ind}}$)~\cite{F1982,CYA17,GCR20,Goncalves_SpringerTheses}
\begin{subequations}
\begin{align}
\dper(\omega) &= \frac{\int_{-\infty}^{\infty} dz \, z \, \rho^{\text{ind}}(z,\omega)}{\int_{-\infty}^{\infty} dz\, \rho^{\text{ind}}(z,\omega)},\\[0.5em]
\dpar(\omega) &= \frac{\int_{-\infty}^{\infty} dz\, z\, \frac{\partial}{\partial z} J_x^{\text{ind}} (z,\omega)}{\int_{-\infty}^{\infty} dz\, \frac{\partial}{\partial z} J_x^{\text{ind}}(z,\omega)} ,
\end{align}
\end{subequations}
in the long-wavelength limit. The Feibelman parameters can be rigorously incorporated in electrodynamic problems by appropriately adjusting the boundary conditions at the surface~\cite{F1982}, both in analytical treatments~\cite{Yan:2015,CYA17,GCR20,Goncalves_SpringerTheses} and in numerical implementations~\cite{Yan:2015,YZY19}.
 
It is implicit that in the presence of more mechanisms, $\dper\simeq\sum_j \dper^{\scriptscriptstyle (j)}$ and $\dpar\simeq\sum_j \dpar^{\scriptscriptstyle (j)}$. In the following we separately consider contributions from bulk spatial dispersion---relevant for any metal surface with a compressible electron gas---and Shockley surface states---relevant to the (111) noble-metal surfaces.

\subsubsection{Contributions from bulk spatial dispersion}

The surface-response functions associated with the spatial dispersion (nonlocal response) of the bulk response functions of the metal, i.e., the wave vector dependence of the dielectric function, can be expressed through the specular reflection model\cite{Ford:1984,Pitarke:2007,Goncalves_SpringerTheses}

\begin{subequations}
\begin{align}
    \dper(\omega)&= -\frac{2}{\pi} \frac{\epsm(\omega) \epsd(\omega)}{\epsm(\omega)-\epsd(\omega)} \int_0^\infty \frac{dk}{k^2} \left[ \frac{1}{\eps_{\rm L}(k,\omega)}-\frac{1}{\epsm(\omega)} \right], \\[0.4em]
    \dpar(\omega)&= 0.
\end{align}
\end{subequations}
Note how the vanishing of $\dpar$ is a consequence of the charge-neutral interface~\cite{Liebsch:1997}. Below, we discuss how this is changed in the presence of a Shockley surface state.

\subsubsection{Contributions from Shockley surface states}

We consider the effect of a Shockley surface state, while for simplicity leaving out the response associated with the spatial dispersion of the bulk states. We note that, with our sign convention of the surface-normal, the surface conductivity $\sigma_{\text{2D}}$ is related to the Feibelman $\dpar$ parameter as $\sigma_{\text{2D}}(\omega) = -\ii \omega \ep_0 (\ep_{\text{m}} - \ep_{\text{d}}) \dpar (\omega)$~\cite{Goncalves_SpringerTheses}. Furthermore, in the nonretarded limit, the 2D plasmon dispersion relation associated with $\sigma_{\text{2D}}$ is generally given by $q(\omega) =\ii\omega\ep_0 (\ep_{\text{m}} + \ep_{\text{d}})/ \sigma_{\text{2D}}(\omega)$~\cite{Goncalves_SpringerTheses}. Anticipating that we have a Shockley surface state that supports an acoustic plasmon~\cite{Echenique:2001}, corresponding to a dispersion relation $\sqrt{\omega(\omega+\ii\gamma_{\rm 2D})}= v_\phi q$, we thus find the connection between this phase velocity $v_\phi$ and the Feibelman parameters to be:
\begin{subequations}
\label{eq:d-acoustic}
\begin{align}
 \dper(\omega)&= 0,\\
\dpar (\omega)&=  -\frac{\ep_{\text{m}}(\omega) + \ep_{\text{d}}(\omega)}{\ep_{\text{m}}(\omega) - \ep_{\text{d}}(\omega)} \frac{v_\phi}{\sqrt{\omega(\omega+\ii\gamma_{\rm 2D})}}.
\end{align}
\end{subequations}
For low frequencies ($\omega\ll \omega_{\text{p}}$) this simplifies to $\dpar(\omega)\simeq v_\phi/\sqrt{\omega(\omega+\ii\gamma_{\rm 2D})}$, while it naturally vanishes at high frequencies. When implemented in the mesoscopic boundary conditions for the electrodynamics, these Feibelman parameters---by construction---support an acoustic plasmon with the desired phase velocity. Considering the poles of the scattering coefficients (see eqs~\ref{eq:rt12} and \ref{eq:rt21}) and substituting in eqs~\ref{eq:d-acoustic}, we obtain
\begin{align}
    0&=\epsm+\epsd-(\epsm-\epsd)q(\dper-\dpar)
%\nonumber\\&
=(\epsm+\epsd)\left(1-\frac{v_\phi q}{\sqrt{\omega(\omega+\ii\gamma_{\rm 2D})}}\right)
\end{align}
and we indeed find two decoupled solutions: the "classical" surface plasmon resonance (defined by $\epsm+\epsd=0$) and the (added \emph{ad-hoc}) acoustic one with $\sqrt{\omega(\omega+\ii\gamma_{\rm 2D})}=v_\phi q$. Table \ref{SS_table} gathers the obtained phase velocities by fitting the acoustic surface plasmon featured in the optical response (see Figure~\ref{fig_SLM_graphene}a), and the damping $\gamma_{\rm 2D}$ by associating it to the width of a Lorentzian; specifically, the widths have been computed numerically from the second derivative of the imaginary part of the reflection coefficient in order to remove the background contribution (see Figure~\ref{fig_SLM_graphene}d).

\subsubsection{Extraction of the $d$-parameters for crystalline metal surfaces} 

We describe crystalline metal films quantum mechanically, simulating their optical response at the level of the random-phase approximation (RPA) following the procedure of ref~\citenum{paper329}: As explained therein, metal films are considered to have translational symmetry in the $\Rb=(x,y)$ plane, so that their electronic wave functions are amenable to expansion in a plane wave basis according to $\Psi_{j,\kparb}(\rb)=A^{-1/2} \ee^{\ii \kparb \cdot \Rb} \varphi_j(z)$, with $A$ denoting the normalization area, $\kparb$ the 2D electron momentum, and $\varphi_j(z)$ the spatial dependence of state $j$ in the quantization direction $z$; the latter quantity is obtained by solving the eigenvalue problem $\Hh\varphi_j(z)=\hbar \en_j^\perp \varphi_j(z)$ to obtain the associated energy eigenvalues $\hbar \en_j^\perp$ of the 1D Hamiltonian $\Hh=-\hbar^2\partial_z^2/2m_e + V(z)$ determining the band dispersion $\hbar\en_{j,\kparb}=\hbar^2\kparb^2/2 m_e+\hbar\en_j^\perp$. The 1D potential $V(z)$, here referred to as the \emph{atomic layer potential} (ALP), is selected from those reported in ref~\citenum{CSE99} that characterize faceted metals of thickness $t$ composed of $N$ atomic planes stacked in the $z$-direction with inter-layer spacing $a_s$ (naturally, $t$ is an integer multiple of $a_s$).

Electronic bands are populated by successively filling the lowest bands until the effective bulk electronic density $n_{\rm eff}$ is reached, thereby determining the Fermi energy $\EF$ as
\begin{equation}
    \EF=\left(\sum_{j=1}^Mm^*_j\right)^{-1} \left(n_{\rm eff} t \hbar^2\pi+\sum_{j=1}^M m^*_j\en_j^\perp \right),
    \label{eq:EF}
\end{equation}
where the sums over $j$ terminate when $\en_M^\perp<\EF/\hbar<\en_{M+1}^\perp$, i.e., $j=M$ is the highest partially-occupied band. The electronic densities $n_{\rm eff}$ are determined by imposing the experimentally-established value of $\EF$ for a given noble metal in the bulk limit (i.e., for a sufficiently thick film). For consistency with experimental observations~\cite{MMC90,MTH04,BCB06}, we impose a linear variation in the effective mass of the parabolic bands as a function of their quantized energies $\hbar\en_j^\perp$ according to $m^*_j/\me=a\hbar\en_j^\perp+b$, thereby avoiding artifacts due to an unrealistic number of excitation channels for vertical transitions introduced by perfectly-aligned parabolic bands; specific parameters used in our calculations are reported in Table\ \ref{bands_paramters}---note that the surface states for (111) noble metal facets are assigned specific experimentally-determined effective masses.

We characterize the optical response of noble metal films by the reflection coefficient $R(\Qb,\omega)$, expressed as a function of the optical in-plane wave vector $\Qb$ and frequency $\omega$. Considering that the relevant length scales are far smaller than the involved optical wavelengths, we invoke the quasistatic approximation to compute the reflection coefficient in terms of electrostatic potentials as $R=1-\phi(z)/\phi^{\rm ext}(z)$, where $\phi=\phi^{\rm ext}+\phi^{\rm ind}$ is the sum of external and induced potentials, the former exciting the system and the latter obtained as $\phiind(z) = \int dz' \chi(z,z')\phiext(z')$ in terms of the film susceptibility $\chi$, which we compute in the ALP-RPA formalism reported in ref~\citenum{paper329}. In principle, the RPA response function is constructed by summing over all possible transitions between electronic states; however, because the Shockley surface state of the (111) facet is incorporated in $\dpar$ following the \emph{ad-hoc} prescription of the previous section, the reflection coefficient that is used to extract $\dper$ is computed by excluding intraband transitions involving surface states, thereby avoiding double-counting of the associated 2DEG.

Following the RPA description of ref~\citenum{paper329}, we correct the Coulomb interaction to incorporate screening from core electrons using the experimentally-extracted polycrystalline dielectric functions $\epsb(\omega)$ parametrized in Table\ \ref{Drude_table} and plotted below in Fig.\ \ref{figS1}. The $\dper$ associated with a given noble metal facet is extracted from ALP-RPA simulations of a sufficiently thick film, so that the optical response is converged with the number of atomic planes. More specifically, we obtain $\dper$ by fitting eq\ \ref{eq:rt12} to the ALP-RPA reflection coefficient of the thick film, employing the corresponding bulk dielectric function $\epsm$ of eq\ \ref{eq:Drude}; importantly, the bulk plasma frequency $\wp$ for each metal facet that enters eq\ \ref{eq:Drude} in the fitting is obtained from the ALP-RPA response of a finite film for sufficiently small in-plane wave vector, e.g., $Q\sim 0.005$, so that nonlocal effects are safely neglected and the surface plasmon resonance ($\sim 1$\,eV) is captured in an uncorrected Fabry--P\'erot description; this procedure enables a stable parametrization of a crystallographic facet's bulk properties when the contribution from the $d$-parameters is negligible. In practice, the surface plasmon for a finite film with $N=10-40$ atomic planes appears at lower energies than the surface plasmon for the semi-infinite film, and the associated resonance is undamped by interband transitions, thereby giving rise to a well-defined peak (see, for instance, Figure~\ref{fig_thin_films}) from which $\wp$ is obtained by fitting $\epsm = \epsd(1+\rdm)/(1-\rdm)$, c.f. eq\ \ref{eq:rt12} in the $Q\to0$ limit.

\begin{figure}[h!]
\centering
\includegraphics[width=0.6\textwidth]{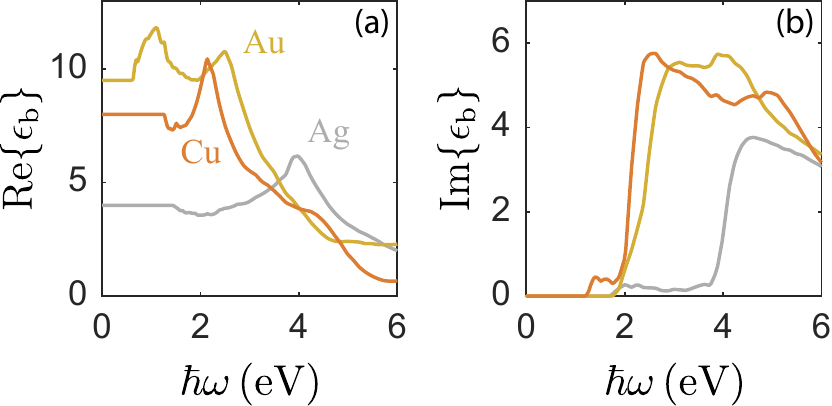}
\caption{{\bf Background dielectric function.} (a) real and (b) imaginary part of $\epsb$ for the noble metals under consideration, obtained by subtracting $-\wp^2/\omega(\omega+\ii \gamma^{\rm exp})$ from the experimental tabulated data of ref~\citenum{JC1972}; see Table~\ref{bands_paramters} for the characteristic parameters of each metal.}
\label{figS1} 
\end{figure}

Once the bulk properties for each facet are set, we construct $\dper(\omega)$ at a given $\omega$ by fitting eq~\ref{eq:FP_equation} for a given value of $Q$ to the ALP-RPA response. We maintain a large number of layers to avoid quantum finite size effects emerging in thin films ($<20$ atomic layers). We then confirm that convergence is maintained with the calculated $\dper(\omega)$ as the number of layers and/or parallel wave vector is varied, where the latter condition is typically satisfied for $0.3<Q<1.5$ \nm. After the parameters are obtained, they are applied to thick (Figure~\ref{fig_SP}) and thin films (Figure~\ref{fig_thin_films}) to confirm their applicability.

In the case of the (100)-facet metals we set $\dpar=0$, whereas for the (111)-facet the prescription of eq~\ref{eq:d-acoustic} is employed to describe the intrinsic low-energy acoustic plasmon; as anticipated, $\dper(\omega)$ approaches zero when $\omega\to0$, and hence such a condition was imposed after solving for $\dper(\omega)$.

\begin{table*}
\centering
\begin{tabular}{c|cccccccc}  \hline
Material & $a (\rm eV^{-1}$) & $b$ & $m^*$(SS)$/m_e$ & $m_0/m_e$& $n_{\rm eff}/n_0$ & $\EF$ (eV) & $\hbar\wp$ (eV) & $\hbar\gamma^{\rm exp}$ (eV)  \\ \hline
Ag(100) &  -0.0817  &  0.2116 &         -         & 0.40~\cite{GPC03} & 0.8710 & -4.43~\cite{CSE99}        & 8.80  & 0.021 \\ 
Ag(111) &  -0.1549  & -0.5446 & 0.40~\cite{RNS01} & 0.25~\cite{SPC05} & 0.8381 & -4.63~\cite{KG1987,PMM95} & 9.19  & 0.021 \\ \hline
Au(100) &  -0.1068  & -0.1802 &         -         & 0.24              & 0.9199 & -5.47~\cite{M1977}        & 8.67  & 0.071 \\ 
Au(111) &  -0.1660  & -0.8937 & 0.26~\cite{RNS01} & 0.26~\cite{SPC05} & 0.9443 & -5.50~\cite{KG1987,PMM95} & 9.88  & 0.071 \\ \hline
Cu(100) &  -0.0751  &  0.1078 &         -         & 0.34              & 0.9634 & -4.59~\cite{M1977}        & 11.38 & 0.103 \\ 
Cu(111) &  -0.1084  & -0.3303 & 0.41~\cite{RNS01} & 0.31~\cite{SPC05} & 0.9285 & -4.98~\cite{M1977}        & 11.50 & 0.103 \\ \hline     
\end{tabular}
\caption{{\bf Characterization of quantum well states in noble metals.} The parameters defining the electronic bands of noble metals entering our optical response calculations are presented. The quantities $a$ and $b$ define the linear variation in effective mass for band $j$ as a function of its associated energy $\hbar\en_j^\perp$; values for the Fermi energy $\EF$, the effective mass associated with surface states $m^*$(SS), and the effective mass for the bottom of the conduction band $m_0$, are extracted from experimental reports, while the effective electron density $n_{\rm eff}$ is fitted to match $\EF$, thus fixing the plasma frequency $\omega_{\rm p}$.}
\label{bands_paramters}
\end{table*}

\subsection{Finite-element implementation}

To calculate the extinction spectra of the truncated octahedra, we used the commercial FEM solver Comsol Multiphysics 5.4. As it has been shown elsewhere~\cite{YZY19,GCR20,Yan:2015}, Feibelman parameters can be incorporated in any computational method by adjusting the boundary conditions. More specifically, the Feibelman parameters introduce discontinuities in the parallel components of the electric and magnetic fields, which can be expressed through~\cite{YZY19,GCR20,Yan:2015}
\begin{subequations}
\begin{equation}\label{Eq:Boundary1}
\hat{\mathbf{n}} \times \left(\mathbf{E}_{2} - \mathbf{E}_{1}\right) =
-\dper \hat{\mathbf{n}} \times \left[\nabla_{\parallel} \hat{\mathbf{n}}
\cdot \left(\mathbf{E}_{2} - \mathbf{E}_{1}\right) \right]~,
\end{equation}
\begin{equation}\label{Eq:Boundary2}
\hat{\mathbf{n}} \times \left(\mathbf{H}_{2} - \mathbf{H}_{1}\right) =
\mathrm{i} \omega \dpar \hat{\mathbf{n}} \times \left[
\nabla_{\parallel} \hat{\mathbf{n}}
\cdot \left(\mathbf{D}_{2} - \mathbf{D}_{1}\right) 
\times \hat{\mathbf{n}}\right]~,
\end{equation}
\end{subequations}
where $\mathbf{E_{i}}$, $\mathbf{H}_{i}$, and $\mathbf{D}_{i}$ are the electric, magnetic, and displacement fields on side $i$ of an interface between two media $1$ and $2$, and $\hat{\mathbf{n}}$ is the unit vector normal to the interface. These conditions can be readily implemented in version 5.4 of Comsol Multiphysics, through surface current and surface magnetic current densities, expressed through the down and up functions in Comsol Multiphysics for the fields at sides 1 and 2. Since the expressions for these currents (right-hand sides of eqs~\ref{Eq:Boundary1} and \ref{Eq:Boundary2}) contain the fields themselves, the problem needs to be solved iteratively, starting with the currents due to the incident plane wave. To calculate scattering and absorption cross section, we need to integrate the Poynting flux of the scattered and total field over a surface (a sphere) enclosing the NP, with radius large enough ($2-3$\;nm more than $R_{\mathrm{circ}}$) to ensure that numerical noise due to the currents close to the surface will be minimum. For sharp-edged NPs like the octahedra studied here, it is also necessary to introduce some rounding, to ensure that any spurious edge/corner modes will be absent. This is needed in the classical calculation, in the absence of surface currents; when these are present, the additional damping they introduce smooths things nicely. However, for a direct comparison between the two cases, it is necessary to include the same rounding in both cases.
This, however, causes an additional numerical problem, because the iterative method diverges when surface currents are added in such small rounded elements. For this reason, surface currents are only used to describe the square $(100)$ and hexagonal $(111)$ facets. This is in practice not a bad approximation, as one needs to somehow introduce a smooth transition between the two different current densities.
In terms of set-up parameters, a cubic physical domain of side $300$\;nm was used, surrounded by $300$\;nm-thick perfectly-matched layers. For the finite-element discretization, a mesh of 30000 domain elements with maximum element size $20$\;nm and minimum element size $0.5$\;nm provided converged spectra.

% =================================================
% --------------------   //    --------------------
% =================================================
\twocolumn

\begin{acknowledgement}
A.~R.~E. and F.~J.~G.~A. acknowledge support from ERC (Advanced Grant 789104-eNANO), the Spanish MINECO (MAT2017-88492-R and SEV2015-0522), the Catalan CERCA Program, and Fundaci\'o Privada Cellex.
N.~A.~M. is a VILLUM Investigator supported by VILLUM FONDEN (grant No. 16498) and Independent Research Fund Denmark (grant no. 7026-00117B).
The Center for Nano Optics is financially supported by the University
of Southern Denmark (SDU 2020 funding).
Simulations were supported by the DeIC National HPC Centre, SDU.
\end{acknowledgement}

%\begin{suppinfo}

%\end{suppinfo}

\providecommand{\latin}[1]{#1}
\makeatletter
\providecommand{\doi}
  {\begingroup\let\do\@makeother\dospecials
  \catcode`\{=1 \catcode`\}=2 \doi@aux}
\providecommand{\doi@aux}[1]{\endgroup\texttt{#1}}
\makeatother
\providecommand*\mcitethebibliography{\thebibliography}
\csname @ifundefined\endcsname{endmcitethebibliography}
  {\let\endmcitethebibliography\endthebibliography}{}


\begin{mcitethebibliography}{81}
\providecommand*\natexlab[1]{#1}
\providecommand*\mciteSetBstSublistMode[1]{}
\providecommand*\mciteSetBstMaxWidthForm[2]{}
\providecommand*\mciteBstWouldAddEndPuncttrue
  {\def\EndOfBibitem{\unskip.}}
\providecommand*\mciteBstWouldAddEndPunctfalse
  {\let\EndOfBibitem\relax}
\providecommand*\mciteSetBstMidEndSepPunct[3]{}
\providecommand*\mciteSetBstSublistLabelBeginEnd[3]{}
\providecommand*\EndOfBibitem{}
\mciteSetBstSublistMode{f}
\mciteSetBstMaxWidthForm{subitem}{(\alph{mcitesubitemcount})}
\mciteSetBstSublistLabelBeginEnd
  {\mcitemaxwidthsubitemform\space}
  {\relax}
  {\relax}

\bibitem[Gramotnev and Bozhevolnyi(2010)Gramotnev, and
  Bozhevolnyi]{Gramotnev:2010}
Gramotnev,~D.~K.; Bozhevolnyi,~S.~I. Plasmonics beyond the diffraction limit.
  \emph{Nat. Photon.} \textbf{2010}, \emph{4}, 83--91\relax
\mciteBstWouldAddEndPuncttrue
\mciteSetBstMidEndSepPunct{\mcitedefaultmidpunct}
{\mcitedefaultendpunct}{\mcitedefaultseppunct}\relax
\EndOfBibitem
\bibitem[Gramotnev and Bozhevolnyi(2014)Gramotnev, and
  Bozhevolnyi]{Gramotnev:2014}
Gramotnev,~D.~K.; Bozhevolnyi,~S.~I. Nanofocusing of electromagnetic radiation.
  \emph{Nat. Photon.} \textbf{2014}, \emph{8}, 14--23\relax
\mciteBstWouldAddEndPuncttrue
\mciteSetBstMidEndSepPunct{\mcitedefaultmidpunct}
{\mcitedefaultendpunct}{\mcitedefaultseppunct}\relax
\EndOfBibitem
\bibitem[Fern{\'a}ndez-Dom{\'i}nguez
  \latin{et~al.}(2017)Fern{\'a}ndez-Dom{\'i}nguez, Garc{\'i}a-Vidal, and
  Mart{\'i}n-Moreno]{Fernandez-Dominguez:2017}
Fern{\'a}ndez-Dom{\'i}nguez,~A.~I.; Garc{\'i}a-Vidal,~F.~J.;
  Mart{\'i}n-Moreno,~L. Unrelenting plasmons. \emph{Nat. Photon.}
  \textbf{2017}, \emph{11}, 8--10\relax
\mciteBstWouldAddEndPuncttrue
\mciteSetBstMidEndSepPunct{\mcitedefaultmidpunct}
{\mcitedefaultendpunct}{\mcitedefaultseppunct}\relax
\EndOfBibitem
\bibitem[Lal \latin{et~al.}(2008)Lal, Clare, and Halas]{Lal:2008}
Lal,~S.; Clare,~S.~E.; Halas,~N.~J. Nanoshell-Enabled Photothermal Cancer
  Therapy: Impending Clinical Impact. \emph{Acc. Chem. Res.} \textbf{2008},
  \emph{41}, 1842--1851\relax
\mciteBstWouldAddEndPuncttrue
\mciteSetBstMidEndSepPunct{\mcitedefaultmidpunct}
{\mcitedefaultendpunct}{\mcitedefaultseppunct}\relax
\EndOfBibitem
\bibitem[Rastinehad \latin{et~al.}(2019)Rastinehad, Anastos, Wajswol, Winoker,
  Sfakianos, Doppalapudi, Carrick, Knauer, Taouli, Lewis, Tewari, Schwartz,
  Canfield, George, West, and Halas]{Rastinehad:2019}
Rastinehad,~A.~R. \latin{et~al.}  Gold nanoshell-localized photothermal
  ablation of prostate tumors in a clinical pilot device study. \emph{Proc.
  Natl. Acad. Sci. U. S. A.} \textbf{2019}, \emph{116}, 18590--18596\relax
\mciteBstWouldAddEndPuncttrue
\mciteSetBstMidEndSepPunct{\mcitedefaultmidpunct}
{\mcitedefaultendpunct}{\mcitedefaultseppunct}\relax
\EndOfBibitem
\bibitem[Wang \latin{et~al.}(2012)Wang, Huang, Dai, and Whangbo]{WHD12}
Wang,~P.; Huang,~B.; Dai,~Y.; Whangbo,~M.-H. Plasmonic photocatalysts:
  harvesting visible light with noble metal nanoparticles. \emph{Phys. Chem.
  Chem. Phys.} \textbf{2012}, \emph{14}, 9813--9825\relax
\mciteBstWouldAddEndPuncttrue
\mciteSetBstMidEndSepPunct{\mcitedefaultmidpunct}
{\mcitedefaultendpunct}{\mcitedefaultseppunct}\relax
\EndOfBibitem
\bibitem[Zhang \latin{et~al.}(2013)Zhang, Chen, Liu, and Tsai]{Zhang:2013}
Zhang,~X.; Chen,~Y.~L.; Liu,~R.-S.; Tsai,~D.~P. Plasmonic photocatalysis.
  \emph{Rep. Prog. Phys.} \textbf{2013}, \emph{76}, 046401\relax
\mciteBstWouldAddEndPuncttrue
\mciteSetBstMidEndSepPunct{\mcitedefaultmidpunct}
{\mcitedefaultendpunct}{\mcitedefaultseppunct}\relax
\EndOfBibitem
\bibitem[Kristensen \latin{et~al.}(2017)Kristensen, Yang, Bozhevolnyi, Link,
  Nordlander, Halas, and Mortensen]{Kristensen:2017}
Kristensen,~A.; Yang,~J. K.~W.; Bozhevolnyi,~S.~I.; Link,~S.; Nordlander,~P.;
  Halas,~N.~J.; Mortensen,~N.~A. Plasmonic colour generation. \emph{Nat. Rev.
  Mater.} \textbf{2017}, \emph{2}, 16088\relax
\mciteBstWouldAddEndPuncttrue
\mciteSetBstMidEndSepPunct{\mcitedefaultmidpunct}
{\mcitedefaultendpunct}{\mcitedefaultseppunct}\relax
\EndOfBibitem
\bibitem[Atwater and Polman(2010)Atwater, and Polman]{Atwater:2010}
Atwater,~H.~A.; Polman,~A. Plasmonics for improved photovoltaic devices.
  \emph{Nat. Mater.} \textbf{2010}, \emph{9}, 205--213\relax
\mciteBstWouldAddEndPuncttrue
\mciteSetBstMidEndSepPunct{\mcitedefaultmidpunct}
{\mcitedefaultendpunct}{\mcitedefaultseppunct}\relax
\EndOfBibitem
\bibitem[Smith \latin{et~al.}(2015)Smith, Faucheaux, and Jain]{SFJ15}
Smith,~J.~G.; Faucheaux,~J.~A.; Jain,~P.~K. Plasmon resonances for solar energy
  harvesting: a mechanistic outlook. \emph{Nano Today} \textbf{2015},
  \emph{10}, 67--80\relax
\mciteBstWouldAddEndPuncttrue
\mciteSetBstMidEndSepPunct{\mcitedefaultmidpunct}
{\mcitedefaultendpunct}{\mcitedefaultseppunct}\relax
\EndOfBibitem
\bibitem[Boltasseva and Atwater(2011)Boltasseva, and Atwater]{BA11}
Boltasseva,~A.; Atwater,~H.~A. Low-loss plasmonic metamaterials. \emph{Science}
  \textbf{2011}, \emph{331}, 290--291\relax
\mciteBstWouldAddEndPuncttrue
\mciteSetBstMidEndSepPunct{\mcitedefaultmidpunct}
{\mcitedefaultendpunct}{\mcitedefaultseppunct}\relax
\EndOfBibitem
\bibitem[{Alonso Calafell} \latin{et~al.}(2019){Alonso Calafell}, Cox,
  Radonji{\'c}, Saavedra, {Garc\'{\i}a de Abajo}, Rozema, and
  Walther]{paper327}
{Alonso Calafell},~I.; Cox,~J.~D.; Radonji{\'c},~M.; Saavedra,~J. R.~M.;
  {Garc\'{\i}a de Abajo},~F.~J.; Rozema,~L.~A.; Walther,~P. Quantum computing
  with graphene plasmons. \emph{npj\ Quantum\ Inf.} \textbf{2019}, \emph{5},
  37\relax
\mciteBstWouldAddEndPuncttrue
\mciteSetBstMidEndSepPunct{\mcitedefaultmidpunct}
{\mcitedefaultendpunct}{\mcitedefaultseppunct}\relax
\EndOfBibitem
\bibitem[Fern{\'a}ndez-Dom{\'i}nguez
  \latin{et~al.}(2018)Fern{\'a}ndez-Dom{\'i}nguez, Bozhevolnyi, and
  Mortensen]{Fernandez-Dominguez:2018}
Fern{\'a}ndez-Dom{\'i}nguez,~A.~I.; Bozhevolnyi,~S.~I.; Mortensen,~N.~A.
  Plasmon-Enhanced Generation of Nonclassical Light. \emph{ACS Photonics}
  \textbf{2018}, \emph{5}, 3447--3451\relax
\mciteBstWouldAddEndPuncttrue
\mciteSetBstMidEndSepPunct{\mcitedefaultmidpunct}
{\mcitedefaultendpunct}{\mcitedefaultseppunct}\relax
\EndOfBibitem
\bibitem[Zhou \latin{et~al.}(2016)Zhou, Zhao, Yu, Ai, Möhwald, Chiechi, Yang,
  and Zhang]{Zhou:2016}
Zhou,~Z.; Zhao,~Z.; Yu,~Y.; Ai,~B.; Möhwald,~H.; Chiechi,~R.~C.; Yang,~J.
  K.~W.; Zhang,~G. From 1D to 3D: Tunable Sub-10 nm Gaps in Large Area Devices.
  \emph{Adv. Mater.} \textbf{2016}, \emph{28}, 2956--2963\relax
\mciteBstWouldAddEndPuncttrue
\mciteSetBstMidEndSepPunct{\mcitedefaultmidpunct}
{\mcitedefaultendpunct}{\mcitedefaultseppunct}\relax
\EndOfBibitem
\bibitem[Baumberg \latin{et~al.}(2019)Baumberg, Aizpurua, Mikkelsen, and
  Smith]{Baumberg:2019}
Baumberg,~J.~J.; Aizpurua,~J.; Mikkelsen,~M.~H.; Smith,~D.~R. Extreme
  nanophotonics from ultrathin metallic gaps. \emph{Nat. Mater.} \textbf{2019},
  \emph{18}, 668--678\relax
\mciteBstWouldAddEndPuncttrue
\mciteSetBstMidEndSepPunct{\mcitedefaultmidpunct}
{\mcitedefaultendpunct}{\mcitedefaultseppunct}\relax
\EndOfBibitem
\bibitem[Kossoy \latin{et~al.}(2014)Kossoy, Merk, Simakov, Leosson,
  K\'ena-Cohen, and Maier]{KMS14}
Kossoy,~A.; Merk,~V.; Simakov,~D.; Leosson,~K.; K\'ena-Cohen,~S.; Maier,~S.~A.
  Optical and Structural Properties of Ultra-thin Gold Films. \emph{Adv.\ Opt.\
  Mater.} \textbf{2014}, \emph{3}, 71--77\relax
\mciteBstWouldAddEndPuncttrue
\mciteSetBstMidEndSepPunct{\mcitedefaultmidpunct}
{\mcitedefaultendpunct}{\mcitedefaultseppunct}\relax
\EndOfBibitem
\bibitem[Maniyara \latin{et~al.}(2019)Maniyara, Rodrigo, Yu, Canet-Ferrer,
  Ghosh, Yongsunthon, Baker, Rezikyan, {Garc\'{\i}a de Abajo}, and
  Pruneri]{paper326}
Maniyara,~R.~A.; Rodrigo,~D.; Yu,~R.; Canet-Ferrer,~J.; Ghosh,~D.~S.;
  Yongsunthon,~R.; Baker,~D.~E.; Rezikyan,~A.; {Garc\'{\i}a de Abajo},~F.~J.;
  Pruneri,~V. Tunable plasmons in ultrathin metal films. \emph{Nat.\ Photon.}
  \textbf{2019}, \emph{13}, 328--333\relax
\mciteBstWouldAddEndPuncttrue
\mciteSetBstMidEndSepPunct{\mcitedefaultmidpunct}
{\mcitedefaultendpunct}{\mcitedefaultseppunct}\relax
\EndOfBibitem
\bibitem[{Abd El-Fattah} \latin{et~al.}(2019){Abd El-Fattah}, Mkhitaryan,
  Brede, Fern\'andez, Li, Guo, Ghosh, {Rodr\'{\i}guez Echarri}, Naveh, Xia,
  Ortega, and {Garc\'{\i}a de Abajo}]{paper335}
{Abd El-Fattah},~Z.~M.; Mkhitaryan,~V.; Brede,~J.; Fern\'andez,~L.; Li,~C.;
  Guo,~Q.; Ghosh,~A.; {Rodr\'{\i}guez Echarri},~A.; Naveh,~D.; Xia,~F.;
  Ortega,~J.~E.; {Garc\'{\i}a de Abajo},~F.~J. Plasmonics in atomically thin
  crystalline silver films. \emph{ACS Nano} \textbf{2019}, \emph{13},
  7771--7779\relax
\mciteBstWouldAddEndPuncttrue
\mciteSetBstMidEndSepPunct{\mcitedefaultmidpunct}
{\mcitedefaultendpunct}{\mcitedefaultseppunct}\relax
\EndOfBibitem
\bibitem[Boltasseva and Shalaev(2019)Boltasseva, and Shalaev]{BS19}
Boltasseva,~A.; Shalaev,~V.~M. Transdimensional Photonics. \emph{ACS\
  Photonics} \textbf{2019}, \emph{6}, 1--3\relax
\mciteBstWouldAddEndPuncttrue
\mciteSetBstMidEndSepPunct{\mcitedefaultmidpunct}
{\mcitedefaultendpunct}{\mcitedefaultseppunct}\relax
\EndOfBibitem
\bibitem[Huang \latin{et~al.}(2010)Huang, Callegari, Geisler, Bruening, Kern,
  Prangsma, Wu, Feichtner, Ziegler, Weinmann, Kamp, Forchel, Biagioni,
  Sennhauser, and Hecht]{Huang:2010}
Huang,~J.-S.; Callegari,~V.; Geisler,~P.; Bruening,~C.; Kern,~J.;
  Prangsma,~J.~C.; Wu,~X.; Feichtner,~T.; Ziegler,~J.; Weinmann,~P.; Kamp,~M.;
  Forchel,~A.; Biagioni,~P.; Sennhauser,~U.; Hecht,~B. Atomically flat
  single-crystalline gold nanostructures for plasmonic nanocircuitry.
  \emph{Nat. Commun.} \textbf{2010}, \emph{1}, 150\relax
\mciteBstWouldAddEndPuncttrue
\mciteSetBstMidEndSepPunct{\mcitedefaultmidpunct}
{\mcitedefaultendpunct}{\mcitedefaultseppunct}\relax
\EndOfBibitem
\bibitem[Hoffmann \latin{et~al.}(2016)Hoffmann, Bashouti, Feichtner,
  Ma\v{c}kovi{\'c}, Dieker, Salaheldin, Richter, Gordan, Zahn, Spiecker, and
  Christiansen]{Hoffmann:2016}
Hoffmann,~B.; Bashouti,~M.~Y.; Feichtner,~T.; Ma\v{c}kovi{\'c},~M.; Dieker,~C.;
  Salaheldin,~A.~M.; Richter,~P.; Gordan,~O.~D.; Zahn,~D. R.~T.; Spiecker,~E.;
  Christiansen,~S. New insights into colloidal gold flakes: structural
  investigation, micro-ellipsometry and thinning procedure towards ultrathin
  monocrystalline layers. \emph{Nanoscale} \textbf{2016}, \emph{8},
  4529--4536\relax
\mciteBstWouldAddEndPuncttrue
\mciteSetBstMidEndSepPunct{\mcitedefaultmidpunct}
{\mcitedefaultendpunct}{\mcitedefaultseppunct}\relax
\EndOfBibitem
\bibitem[M\'{e}jard \latin{et~al.}(2017)M\'{e}jard, Verdy, Demichel, Petit,
  Markey, Herbst, Chassagnon, des Francs, Cluzel, and Bouhelier]{Mejard:2017}
M\'{e}jard,~R.; Verdy,~A.; Demichel,~O.; Petit,~M.; Markey,~L.; Herbst,~F.;
  Chassagnon,~R.; des Francs,~G.~C.; Cluzel,~B.; Bouhelier,~A. Advanced
  engineering of single-crystal gold nanoantennas. \emph{Opt. Mater. Express}
  \textbf{2017}, \emph{7}, 1157--1168\relax
\mciteBstWouldAddEndPuncttrue
\mciteSetBstMidEndSepPunct{\mcitedefaultmidpunct}
{\mcitedefaultendpunct}{\mcitedefaultseppunct}\relax
\EndOfBibitem
\bibitem[Cheng \latin{et~al.}(2019)Cheng, Lee, Choi, Wang, Zhang, Zhang, Gwo,
  Chang, Li, and Shih]{Cheng:2019}
Cheng,~F.; Lee,~C.-J.; Choi,~J.; Wang,~C.-Y.; Zhang,~Q.; Zhang,~H.; Gwo,~S.;
  Chang,~W.-H.; Li,~X.; Shih,~C.-K. Epitaxial Growth of Optically Thick, Single
  Crystalline Silver Films for Plasmonics. \emph{ACS Appl. Mater. Interfaces}
  \textbf{2019}, \emph{11}, 3189--3195\relax
\mciteBstWouldAddEndPuncttrue
\mciteSetBstMidEndSepPunct{\mcitedefaultmidpunct}
{\mcitedefaultendpunct}{\mcitedefaultseppunct}\relax
\EndOfBibitem
\bibitem[McPeak \latin{et~al.}(2015)McPeak, Jayanti, Kress, Meyer, Iotti,
  Rossinelli, and Norris]{MJK15}
McPeak,~K.~M.; Jayanti,~S.~V.; Kress,~S. J.~P.; Meyer,~S.; Iotti,~S.;
  Rossinelli,~A.; Norris,~D.~J. Plasmonic films can easily be better: rules and
  recipes. \emph{ACS Photonics} \textbf{2015}, \emph{2}, 326--333\relax
\mciteBstWouldAddEndPuncttrue
\mciteSetBstMidEndSepPunct{\mcitedefaultmidpunct}
{\mcitedefaultendpunct}{\mcitedefaultseppunct}\relax
\EndOfBibitem
\bibitem[Inglesfield(1982)]{Inglesfield:1982}
Inglesfield,~J.~E. Surface electronic structure. \emph{Rep. Prog. Phys.}
  \textbf{1982}, \emph{45}, 223--284\relax
\mciteBstWouldAddEndPuncttrue
\mciteSetBstMidEndSepPunct{\mcitedefaultmidpunct}
{\mcitedefaultendpunct}{\mcitedefaultseppunct}\relax
\EndOfBibitem
\bibitem[Shockley(1939)]{Shockley:1939}
Shockley,~W. On the Surface States Associated with a Periodic Potential.
  \emph{Phys. Rev.} \textbf{1939}, \emph{56}, 317--323\relax
\mciteBstWouldAddEndPuncttrue
\mciteSetBstMidEndSepPunct{\mcitedefaultmidpunct}
{\mcitedefaultendpunct}{\mcitedefaultseppunct}\relax
\EndOfBibitem
\bibitem[Echenique \latin{et~al.}(2001)Echenique, Osma, Machado, Silkin,
  Chulkov, and Pitarke]{Echenique:2001}
Echenique,~P.~M.; Osma,~J.; Machado,~M.; Silkin,~V.~M.; Chulkov,~E.~V.;
  Pitarke,~J.~M. Surface-state electron dynamics in noble metals. \emph{Prog.
  Surf. Sci.} \textbf{2001}, \emph{67}, 271--283\relax
\mciteBstWouldAddEndPuncttrue
\mciteSetBstMidEndSepPunct{\mcitedefaultmidpunct}
{\mcitedefaultendpunct}{\mcitedefaultseppunct}\relax
\EndOfBibitem
\bibitem[Suto \latin{et~al.}(1989)Suto, Tsuei, Plummer, and
  Burstein]{Suto:1989}
Suto,~S.; Tsuei,~K.-D.; Plummer,~E.~W.; Burstein,~E. Surface-plasmon energy and
  dispersion on Ag single crystals. \emph{Phys. Rev. Lett.} \textbf{1989},
  \emph{63}, 2590--2593\relax
\mciteBstWouldAddEndPuncttrue
\mciteSetBstMidEndSepPunct{\mcitedefaultmidpunct}
{\mcitedefaultendpunct}{\mcitedefaultseppunct}\relax
\EndOfBibitem
\bibitem[Diaconescu \latin{et~al.}(2007)Diaconescu, Pohl, Vattuone, Savio,
  Hofmann, Silkin, Pitarke, Chulkov, Echenique, Far{\'i}as, and
  Rocca]{Diaconescu:2007}
Diaconescu,~B.; Pohl,~K.; Vattuone,~L.; Savio,~L.; Hofmann,~P.; Silkin,~V.~M.;
  Pitarke,~J.~M.; Chulkov,~E.~V.; Echenique,~P.~M.; Far{\'i}as,~D.; Rocca,~M.
  Low-energy acoustic plasmons at metal surfaces. \emph{Nature} \textbf{2007},
  \emph{448}, 57--59\relax
\mciteBstWouldAddEndPuncttrue
\mciteSetBstMidEndSepPunct{\mcitedefaultmidpunct}
{\mcitedefaultendpunct}{\mcitedefaultseppunct}\relax
\EndOfBibitem
\bibitem[Politano \latin{et~al.}(2015)Politano, Silkin, Nechaev, Vitiello,
  Viti, Aliev, Babanly, Chiarello, Echenique, and Chulkov]{Politano:2015}
Politano,~A.; Silkin,~V.~M.; Nechaev,~I.~A.; Vitiello,~M.~S.; Viti,~L.;
  Aliev,~Z.~S.; Babanly,~M.~B.; Chiarello,~G.; Echenique,~P.~M.; Chulkov,~E.~V.
  Interplay of Surface and Dirac Plasmons in Topological Insulators: The Case
  of {Bi$_2$Se$_3$}. \emph{Phys. Rev. Lett.} \textbf{2015}, \emph{115},
  216802\relax
\mciteBstWouldAddEndPuncttrue
\mciteSetBstMidEndSepPunct{\mcitedefaultmidpunct}
{\mcitedefaultendpunct}{\mcitedefaultseppunct}\relax
\EndOfBibitem
\bibitem[Varas \latin{et~al.}(2016)Varas, Garc{\'i}a-Gonz{\'a}lez, Feist,
  Garc{\'i}a-Vidal, and Rubio]{Varas:2016}
Varas,~A.; Garc{\'i}a-Gonz{\'a}lez,~P.; Feist,~J.; Garc{\'i}a-Vidal,~F.~J.;
  Rubio,~A. Quantum plasmonics: from jellium models to ab initio calculations.
  \emph{Nanophotonics} \textbf{2016}, \emph{5}, 409--426\relax
\mciteBstWouldAddEndPuncttrue
\mciteSetBstMidEndSepPunct{\mcitedefaultmidpunct}
{\mcitedefaultendpunct}{\mcitedefaultseppunct}\relax
\EndOfBibitem
\bibitem[Zhu \latin{et~al.}(2016)Zhu, Esteban, Borisov, Baumberg, Nordlander,
  Lezec, Aizpurua, and Crozier]{Zhu:2016}
Zhu,~W.; Esteban,~R.; Borisov,~A.~G.; Baumberg,~J.~J.; Nordlander,~P.;
  Lezec,~H.~J.; Aizpurua,~J.; Crozier,~K.~B. Quantum mechanical effects in
  plasmonic structures with subnanometre gaps. \emph{Nat. Commun.}
  \textbf{2016}, \emph{7}, 11495\relax
\mciteBstWouldAddEndPuncttrue
\mciteSetBstMidEndSepPunct{\mcitedefaultmidpunct}
{\mcitedefaultendpunct}{\mcitedefaultseppunct}\relax
\EndOfBibitem
\bibitem[Feibelman(1982)]{F1982}
Feibelman,~P.~J. Surface electromagnetic fields. \emph{Prog.\ Surf.\ Sci.}
  \textbf{1982}, \emph{12}, 287--407\relax
\mciteBstWouldAddEndPuncttrue
\mciteSetBstMidEndSepPunct{\mcitedefaultmidpunct}
{\mcitedefaultendpunct}{\mcitedefaultseppunct}\relax
\EndOfBibitem
\bibitem[Christensen \latin{et~al.}(2017)Christensen, Yan, Jauho,
  Solja{\v{c}}i{\'c}, and Mortensen]{CYA17}
Christensen,~T.; Yan,~W.; Jauho,~A.-P.; Solja{\v{c}}i{\'c},~M.;
  Mortensen,~N.~A. Quantum corrections in nanoplasmonics: shape, scale, and
  material. \emph{Phys. Rev. Lett.} \textbf{2017}, \emph{118}, 157402\relax
\mciteBstWouldAddEndPuncttrue
\mciteSetBstMidEndSepPunct{\mcitedefaultmidpunct}
{\mcitedefaultendpunct}{\mcitedefaultseppunct}\relax
\EndOfBibitem
\bibitem[Gon\c{c}alves \latin{et~al.}(2020)Gon\c{c}alves, Christensen, Rivera,
  Jauho, Mortensen, and Solja\v{c}i\'{c}]{GCR20}
Gon\c{c}alves,~P. A.~D.; Christensen,~T.; Rivera,~N.; Jauho,~A.-P.;
  Mortensen,~N.~A.; Solja\v{c}i\'{c},~M. Plasmon-Emitter Interactions at the
  Nanoscale. \emph{Nat.\ Commun.} \textbf{2020}, \emph{11}, 366\relax
\mciteBstWouldAddEndPuncttrue
\mciteSetBstMidEndSepPunct{\mcitedefaultmidpunct}
{\mcitedefaultendpunct}{\mcitedefaultseppunct}\relax
\EndOfBibitem
\bibitem[Gon\c{c}alves(2020)]{Goncalves_SpringerTheses}
Gon\c{c}alves,~P. A.~D. \emph{Plasmonics and Light--Matter Interactions in
  Two-Dimensional Materials and in Metal Nanostructures: Classical and Quantum
  Considerations}; Springer Nature, 2020\relax
\mciteBstWouldAddEndPuncttrue
\mciteSetBstMidEndSepPunct{\mcitedefaultmidpunct}
{\mcitedefaultendpunct}{\mcitedefaultseppunct}\relax
\EndOfBibitem
\bibitem[Liebsch(1997)]{Liebsch:1997}
Liebsch,~A. \emph{Electronic Excitations at Metal Surfaces}; Springer: New
  York, 1997\relax
\mciteBstWouldAddEndPuncttrue
\mciteSetBstMidEndSepPunct{\mcitedefaultmidpunct}
{\mcitedefaultendpunct}{\mcitedefaultseppunct}\relax
\EndOfBibitem
\bibitem[Gon\c{c}alves \latin{et~al.}(2020)Gon\c{c}alves, Christensen, Peres,
  Jauho, Epstein, Koppens, Solja{\v{c}}i{\'c}, and Mortensen]{Goncalves_arxiv}
Gon\c{c}alves,~P. A.~D.; Christensen,~T.; Peres,~N. M.~R.; Jauho,~A.-P.;
  Epstein,~I.; Koppens,~F. H.~L.; Solja{\v{c}}i{\'c},~M.; Mortensen,~N.~A.
  Quantum Surface-Response of Metals Revealed by Acoustic Graphene Plasmons.
  \emph{arXiv:2008.07613} \textbf{2020}, \relax
\mciteBstWouldAddEndPunctfalse
\mciteSetBstMidEndSepPunct{\mcitedefaultmidpunct}
{}{\mcitedefaultseppunct}\relax
\EndOfBibitem
\bibitem[Yan \latin{et~al.}(2015)Yan, Wubs, and Mortensen]{Yan:2015}
Yan,~W.; Wubs,~M.; Mortensen,~N.~A. Projected Dipole Model for Quantum
  Plasmonics. \emph{Phys. Rev. Lett.} \textbf{2015}, \emph{115}, 137403\relax
\mciteBstWouldAddEndPuncttrue
\mciteSetBstMidEndSepPunct{\mcitedefaultmidpunct}
{\mcitedefaultendpunct}{\mcitedefaultseppunct}\relax
\EndOfBibitem
\bibitem[Yang \latin{et~al.}(2019)Yang, Zhu, Yan, Agarwal, Zheng, Joannopoulos,
  Lalanne, Christensen, Berggren, and Solja{\v{c}}i{\'c}]{YZY19}
Yang,~Y.; Zhu,~D.; Yan,~W.; Agarwal,~A.; Zheng,~M.; Joannopoulos,~J.~D.;
  Lalanne,~P.; Christensen,~T.; Berggren,~K.~K.; Solja{\v{c}}i{\'c},~M. A
  general theoretical and experimental framework for nanoscale
  electromagnetism. \emph{Nature} \textbf{2019}, \emph{576}, 248--252\relax
\mciteBstWouldAddEndPuncttrue
\mciteSetBstMidEndSepPunct{\mcitedefaultmidpunct}
{\mcitedefaultendpunct}{\mcitedefaultseppunct}\relax
\EndOfBibitem
\bibitem[{Garc\'{i}a de Abajo}(2008)]{GarciadeAbajo:2008}
{Garc\'{i}a de Abajo},~F.~J. Nonlocal effects in the plasmons of strongly
  interacting nanoparticles, dimers, and waveguides. \emph{J. Phys. Chem. C}
  \textbf{2008}, \emph{112}, 17983--17987\relax
\mciteBstWouldAddEndPuncttrue
\mciteSetBstMidEndSepPunct{\mcitedefaultmidpunct}
{\mcitedefaultendpunct}{\mcitedefaultseppunct}\relax
\EndOfBibitem
\bibitem[Johnson and Christy(1972)Johnson, and Christy]{JC1972}
Johnson,~P.~B.; Christy,~R.~W. Optical constants of the noble metals.
  \emph{Phys.\ Rev.\ B} \textbf{1972}, \emph{6}, 4370--4379\relax
\mciteBstWouldAddEndPuncttrue
\mciteSetBstMidEndSepPunct{\mcitedefaultmidpunct}
{\mcitedefaultendpunct}{\mcitedefaultseppunct}\relax
\EndOfBibitem
\bibitem[Echarri \latin{et~al.}(2019)Echarri, Cox, and {Garc\'{\i}a de
  Abajo}]{paper329}
Echarri,~A.~R.; Cox,~J.~D.; {Garc\'{\i}a de Abajo},~F.~J. Quantum Effects in
  the Acoustic Plasmons of Atomically-Thin Heterostructures. \emph{Optica}
  \textbf{2019}, \emph{6}, 630--641\relax
\mciteBstWouldAddEndPuncttrue
\mciteSetBstMidEndSepPunct{\mcitedefaultmidpunct}
{\mcitedefaultendpunct}{\mcitedefaultseppunct}\relax
\EndOfBibitem
\bibitem[Chulkov \latin{et~al.}(1999)Chulkov, Silkin, and Echenique]{CSE99}
Chulkov,~E.~V.; Silkin,~V.~M.; Echenique,~P.~M. Image potential states on metal
  surfaces: binding energies and wave functions. \emph{Surf.\ Sci.}
  \textbf{1999}, \emph{437}, 330--352\relax
\mciteBstWouldAddEndPuncttrue
\mciteSetBstMidEndSepPunct{\mcitedefaultmidpunct}
{\mcitedefaultendpunct}{\mcitedefaultseppunct}\relax
\EndOfBibitem
\bibitem[Ritchie and Marusak(1966)Ritchie, and Marusak]{RM1966}
Ritchie,~R.~H.; Marusak,~A.~L. The surface plasmon dispersion relation for an
  electron gas. \emph{Surf.\ Sci.} \textbf{1966}, \emph{4}, 234--240\relax
\mciteBstWouldAddEndPuncttrue
\mciteSetBstMidEndSepPunct{\mcitedefaultmidpunct}
{\mcitedefaultendpunct}{\mcitedefaultseppunct}\relax
\EndOfBibitem
\bibitem[Ford and Weber(1984)Ford, and Weber]{Ford:1984}
Ford,~G.~W.; Weber,~W.~H. Electromagnetic interactions of molecules with metal
  surfaces. \emph{Phys. Rep.} \textbf{1984}, \emph{113}, 195--287\relax
\mciteBstWouldAddEndPuncttrue
\mciteSetBstMidEndSepPunct{\mcitedefaultmidpunct}
{\mcitedefaultendpunct}{\mcitedefaultseppunct}\relax
\EndOfBibitem
\bibitem[Pitarke \latin{et~al.}(2007)Pitarke, Silkin, Chulkov, and
  Echenique]{Pitarke:2007}
Pitarke,~J.~M.; Silkin,~V.~M.; Chulkov,~E.~V.; Echenique,~P.~M. Theory of
  surface plasmons and surface plasmon polaritons. \emph{Rep. Prog. Phys.}
  \textbf{2007}, \emph{70}, 1--87\relax
\mciteBstWouldAddEndPuncttrue
\mciteSetBstMidEndSepPunct{\mcitedefaultmidpunct}
{\mcitedefaultendpunct}{\mcitedefaultseppunct}\relax
\EndOfBibitem
\bibitem[Rocca \latin{et~al.}(1995)Rocca, Yibing, de~Mongeot, and
  Valbusa]{RYB95}
Rocca,~M.; Yibing,~L.; de~Mongeot,~F.~B.; Valbusa,~U. Surface plasmon
  dispersion and damping on Ag(111). \emph{Phys.\ Rev.\ B} \textbf{1995},
  \emph{52}, 14 947--14 953\relax
\mciteBstWouldAddEndPuncttrue
\mciteSetBstMidEndSepPunct{\mcitedefaultmidpunct}
{\mcitedefaultendpunct}{\mcitedefaultseppunct}\relax
\EndOfBibitem
\bibitem[Park and Palmer(2009)Park, and Palmer]{PP09}
Park,~S.~J.; Palmer,~R.~E. Plasmon dispersion of the Au (111) surface with and
  without self-assembled monolayers. \emph{Phys.\ Rev.\ Lett.} \textbf{2009},
  \emph{102}, 216805\relax
\mciteBstWouldAddEndPuncttrue
\mciteSetBstMidEndSepPunct{\mcitedefaultmidpunct}
{\mcitedefaultendpunct}{\mcitedefaultseppunct}\relax
\EndOfBibitem
\bibitem[Robusto and Braunstein(1981)Robusto, and Braunstein]{RB1981}
Robusto,~P.~F.; Braunstein,~R. Optical measurements of the surface plasmon of
  copper. \emph{Phys.\ Status Solidi B} \textbf{1981}, \emph{107},
  443--449\relax
\mciteBstWouldAddEndPuncttrue
\mciteSetBstMidEndSepPunct{\mcitedefaultmidpunct}
{\mcitedefaultendpunct}{\mcitedefaultseppunct}\relax
\EndOfBibitem
\bibitem[Pitarke \latin{et~al.}(2004)Pitarke, Nazarov, Silkin, Chulkov,
  Zaremba, and Echenique]{Pitarke:2004}
Pitarke,~J.~M.; Nazarov,~V.~U.; Silkin,~V.~M.; Chulkov,~E.~V.; Zaremba,~E.;
  Echenique,~P.~M. Theory of acoustic surface plasmons. \emph{Phys. Rev. B}
  \textbf{2004}, \emph{70}, 205403\relax
\mciteBstWouldAddEndPuncttrue
\mciteSetBstMidEndSepPunct{\mcitedefaultmidpunct}
{\mcitedefaultendpunct}{\mcitedefaultseppunct}\relax
\EndOfBibitem
\bibitem[Silkin \latin{et~al.}(2005)Silkin, Pitarke, Chulkov, and
  Echenique]{SPC05}
Silkin,~V.~M.; Pitarke,~J.~M.; Chulkov,~E.~V.; Echenique,~P.~M. Acoustic
  surface plasmons in the noble metals Cu, Ag, and Au. \emph{Phys.\ Rev.\ B}
  \textbf{2005}, \emph{72}, 115435\relax
\mciteBstWouldAddEndPuncttrue
\mciteSetBstMidEndSepPunct{\mcitedefaultmidpunct}
{\mcitedefaultendpunct}{\mcitedefaultseppunct}\relax
\EndOfBibitem
\bibitem[Pohl \latin{et~al.}(2010)Pohl, Diaconescu, Vercelli, Vattuone, Silkin,
  Chulkov, Echenique, and Rocca]{Pohl:2010}
Pohl,~K.; Diaconescu,~B.; Vercelli,~G.; Vattuone,~L.; Silkin,~V.~M.;
  Chulkov,~E.~V.; Echenique,~P.~M.; Rocca,~M. Acoustic surface plasmon on
  Cu(111). \emph{{EPL} (Europhysics Letters)} \textbf{2010}, \emph{90},
  57006\relax
\mciteBstWouldAddEndPuncttrue
\mciteSetBstMidEndSepPunct{\mcitedefaultmidpunct}
{\mcitedefaultendpunct}{\mcitedefaultseppunct}\relax
\EndOfBibitem
\bibitem[Yan \latin{et~al.}(2012)Yan, Jacobsen, and Thygesen]{Yan:2012}
Yan,~J.; Jacobsen,~K.~W.; Thygesen,~K.~S. Conventional and acoustic surface
  plasmons on noble metal surfaces: A time-dependent density functional theory
  study. \emph{Phys. Rev. B} \textbf{2012}, \emph{86}, 241404\relax
\mciteBstWouldAddEndPuncttrue
\mciteSetBstMidEndSepPunct{\mcitedefaultmidpunct}
{\mcitedefaultendpunct}{\mcitedefaultseppunct}\relax
\EndOfBibitem
\bibitem[Park and Palmer(2010)Park, and Palmer]{PP10}
Park,~S.~J.; Palmer,~R.~E. Acoustic plasmon on the Au (111) surface.
  \emph{Phys.\ Rev.\ Lett.} \textbf{2010}, \emph{105}, 016801\relax
\mciteBstWouldAddEndPuncttrue
\mciteSetBstMidEndSepPunct{\mcitedefaultmidpunct}
{\mcitedefaultendpunct}{\mcitedefaultseppunct}\relax
\EndOfBibitem
\bibitem[Vattuone \latin{et~al.}(2013)Vattuone, Smerieri, Langer, Tegenkamp,
  Pfn{\"u}r, Silkin, Chulkov, Echenique, and Rocca]{VSL13}
Vattuone,~L.; Smerieri,~M.; Langer,~T.; Tegenkamp,~C.; Pfn{\"u}r,~H.;
  Silkin,~V.~M.; Chulkov,~E.~V.; Echenique,~P.~M.; Rocca,~M. Correlated motion
  of electrons on the Au (111) surface: anomalous acoustic surface-plasmon
  dispersion and single-particle excitations. \emph{Phys.\ Rev.\ Lett.}
  \textbf{2013}, \emph{110}, 127405\relax
\mciteBstWouldAddEndPuncttrue
\mciteSetBstMidEndSepPunct{\mcitedefaultmidpunct}
{\mcitedefaultendpunct}{\mcitedefaultseppunct}\relax
\EndOfBibitem
\bibitem[Reinert \latin{et~al.}(2001)Reinert, Nicolay, Schmidt, Ehm, and
  H{\"u}fner]{RNS01}
Reinert,~F.; Nicolay,~G.; Schmidt,~S.; Ehm,~D.; H{\"u}fner,~S. Direct
  measurements of the L-gap surface states on the (111) face of noble metals by
  photoelectron spectroscopy. \emph{Phys.\ Rev.\ B} \textbf{2001}, \emph{63},
  115415\relax
\mciteBstWouldAddEndPuncttrue
\mciteSetBstMidEndSepPunct{\mcitedefaultmidpunct}
{\mcitedefaultendpunct}{\mcitedefaultseppunct}\relax
\EndOfBibitem
\bibitem[Principi \latin{et~al.}(2018)Principi, {van Loon}, Polini, and
  Katsnelson]{PVP18}
Principi,~A.; {van Loon},~E.; Polini,~M.; Katsnelson,~M.~I. Confining graphene
  plasmons to the ultimate limit. \emph{Phys.\ Rev.\ B} \textbf{2018},
  \emph{98}, 035427\relax
\mciteBstWouldAddEndPuncttrue
\mciteSetBstMidEndSepPunct{\mcitedefaultmidpunct}
{\mcitedefaultendpunct}{\mcitedefaultseppunct}\relax
\EndOfBibitem
\bibitem[{Alcaraz Iranzo} \latin{et~al.}(2018){Alcaraz Iranzo}, Nanot, Dias,
  Epstein, Peng, Efetov, Lundeberg, Parret, Osmond, Hong, Kong, Englund, Peres,
  and Koppens]{AND18}
{Alcaraz Iranzo},~D.; Nanot,~S.; Dias,~E. J.~C.; Epstein,~I.; Peng,~C.;
  Efetov,~D.~K.; Lundeberg,~M.~B.; Parret,~R.; Osmond,~J.; Hong,~J.-Y.;
  Kong,~J.; Englund,~D.~R.; Peres,~N. M.~R.; Koppens,~F. H.~L. Probing the
  ultimate plasmon confinement limits with a van der Waals heterostructure.
  \emph{Science} \textbf{2018}, \emph{360}, 291--295\relax
\mciteBstWouldAddEndPuncttrue
\mciteSetBstMidEndSepPunct{\mcitedefaultmidpunct}
{\mcitedefaultendpunct}{\mcitedefaultseppunct}\relax
\EndOfBibitem
\bibitem[Lee \latin{et~al.}(2019)Lee, Yoo, Avouris, Low, and Oh]{LYA19}
Lee,~I.-H.; Yoo,~D.; Avouris,~P.; Low,~T.; Oh,~S.-H. Graphene acoustic plasmon
  resonator for ultrasensitive infrared spectroscopy. \emph{Nat.\ Nanotechnol.}
  \textbf{2019}, \emph{14}, 313–--319\relax
\mciteBstWouldAddEndPuncttrue
\mciteSetBstMidEndSepPunct{\mcitedefaultmidpunct}
{\mcitedefaultendpunct}{\mcitedefaultseppunct}\relax
\EndOfBibitem
\bibitem[Epstein \latin{et~al.}(2020)Epstein, Alcaraz, Huang, Pusapati,
  Hugonin, Kumar, Deputy, Khodkov, Rappoport, Hong, Peres, Kong, Smith, and
  Koppens]{EAH20}
Epstein,~I.; Alcaraz,~D.; Huang,~Z.; Pusapati,~V.-V.; Hugonin,~J.-P.;
  Kumar,~A.; Deputy,~X.~M.; Khodkov,~T.; Rappoport,~T.~G.; Hong,~J.-Y.;
  Peres,~N. M.~R.; Kong,~J.; Smith,~D.~R.; Koppens,~F. H.~L. Far-field
  excitation of single graphene plasmon cavities with ultracompressed mode
  volumes. \emph{Science} \textbf{2020}, \emph{368}, 1219--1223\relax
\mciteBstWouldAddEndPuncttrue
\mciteSetBstMidEndSepPunct{\mcitedefaultmidpunct}
{\mcitedefaultendpunct}{\mcitedefaultseppunct}\relax
\EndOfBibitem
\bibitem[{Garc\'{\i}a de Abajo}(2014)]{paper235}
{Garc\'{\i}a de Abajo},~F.~J. Graphene Plasmonics: Challenges and
  Opportunities. \emph{ACS\ Photonics} \textbf{2014}, \emph{1}, 135--152\relax
\mciteBstWouldAddEndPuncttrue
\mciteSetBstMidEndSepPunct{\mcitedefaultmidpunct}
{\mcitedefaultendpunct}{\mcitedefaultseppunct}\relax
\EndOfBibitem
\bibitem[Gon\c{c}alves and Peres(2016)Gon\c{c}alves, and Peres]{GoncalvesPeres}
Gon\c{c}alves,~P. A.~D.; Peres,~N. M.~R. \emph{An Introduction to Graphene
  Plasmonics}, 1st ed.; World Scientific: Singapore, 2016\relax
\mciteBstWouldAddEndPuncttrue
\mciteSetBstMidEndSepPunct{\mcitedefaultmidpunct}
{\mcitedefaultendpunct}{\mcitedefaultseppunct}\relax
\EndOfBibitem
\bibitem[Wunsch \latin{et~al.}(2006)Wunsch, Stauber, Sols, and Guinea]{WSS06}
Wunsch,~B.; Stauber,~T.; Sols,~F.; Guinea,~F. Dynamical Polarization of
  Graphene at Finite Doping. \emph{New\ J.\ Phys.} \textbf{2006}, \emph{8},
  318\relax
\mciteBstWouldAddEndPuncttrue
\mciteSetBstMidEndSepPunct{\mcitedefaultmidpunct}
{\mcitedefaultendpunct}{\mcitedefaultseppunct}\relax
\EndOfBibitem
\bibitem[Hwang and Sarma(2007)Hwang, and Sarma]{HS07}
Hwang,~E.~H.; Sarma,~S.~D. Dielectric function, screening, and plasmons in
  two-dimensional graphene. \emph{Phys.\ Rev.\ B} \textbf{2007}, \emph{75},
  205418\relax
\mciteBstWouldAddEndPuncttrue
\mciteSetBstMidEndSepPunct{\mcitedefaultmidpunct}
{\mcitedefaultendpunct}{\mcitedefaultseppunct}\relax
\EndOfBibitem
\bibitem[Mermin(1970)]{M1970}
Mermin,~N.~D. Lindhard dielectric function in the relaxation-time
  approximation. \emph{Phys.\ Rev.\ B} \textbf{1970}, \emph{1},
  2362--2363\relax
\mciteBstWouldAddEndPuncttrue
\mciteSetBstMidEndSepPunct{\mcitedefaultmidpunct}
{\mcitedefaultendpunct}{\mcitedefaultseppunct}\relax
\EndOfBibitem
\bibitem[Myroshnychenko \latin{et~al.}(2008)Myroshnychenko, Carb{\'o}-Argibay,
  Pastoriza-Santos, P{\'e}rez-Juste, Liz-Marz{\'a}n, and Garc{\'i}a~de
  Abajo]{Myroshnychenko:2008a}
Myroshnychenko,~V.; Carb{\'o}-Argibay,~E.; Pastoriza-Santos,~I.;
  P{\'e}rez-Juste,~J.; Liz-Marz{\'a}n,~L.~M.; Garc{\'i}a~de Abajo,~F.~J.
  Modeling the Optical Response of Highly Faceted Metal Nanoparticles with a
  Fully {3D} Boundary Element Method. \emph{Adv. Mater.} \textbf{2008},
  \emph{20}, 4288--4293\relax
\mciteBstWouldAddEndPuncttrue
\mciteSetBstMidEndSepPunct{\mcitedefaultmidpunct}
{\mcitedefaultendpunct}{\mcitedefaultseppunct}\relax
\EndOfBibitem
\bibitem[Mortensen \latin{et~al.}(2014)Mortensen, Raza, Wubs, S{\o}ndergaard,
  and Bozhevolnyi]{Mortensen:2014}
Mortensen,~N.~A.; Raza,~S.; Wubs,~M.; S{\o}ndergaard,~T.; Bozhevolnyi,~S.~I. A
  generalized nonlocal optical response theory for plasmonic nanostructures.
  \emph{Nat. Commun.} \textbf{2014}, \emph{5}, 3809\relax
\mciteBstWouldAddEndPuncttrue
\mciteSetBstMidEndSepPunct{\mcitedefaultmidpunct}
{\mcitedefaultendpunct}{\mcitedefaultseppunct}\relax
\EndOfBibitem
\bibitem[Raza \latin{et~al.}(2015)Raza, Bozhevolnyi, Wubs, and
  Mortensen]{Raza:2015a}
Raza,~S.; Bozhevolnyi,~S.~I.; Wubs,~M.; Mortensen,~N.~A. Nonlocal optical
  response in metallic nanostructures. \emph{J. Phys.: Condens. Matter}
  \textbf{2015}, \emph{27}, 183204\relax
\mciteBstWouldAddEndPuncttrue
\mciteSetBstMidEndSepPunct{\mcitedefaultmidpunct}
{\mcitedefaultendpunct}{\mcitedefaultseppunct}\relax
\EndOfBibitem
\bibitem[Ouyang \latin{et~al.}(1992)Ouyang, Batson, and Isaacson]{Ouyang:1992}
Ouyang,~F.; Batson,~P.~E.; Isaacson,~M. Quantum size effects in the
  surface-plasmon excitation of small metallic particles by
  electron-energy-loss spectroscopy. \emph{Phys. Rev. B} \textbf{1992},
  \emph{46}, 15421--15425\relax
\mciteBstWouldAddEndPuncttrue
\mciteSetBstMidEndSepPunct{\mcitedefaultmidpunct}
{\mcitedefaultendpunct}{\mcitedefaultseppunct}\relax
\EndOfBibitem
\bibitem[Scholl \latin{et~al.}(2012)Scholl, Koh, and Dionne]{Scholl:2012}
Scholl,~J.~A.; Koh,~A.~L.; Dionne,~J.~A. Quantum plasmon resonances of
  individual metallic nanoparticles. \emph{Nature} \textbf{2012}, \emph{483},
  421\relax
\mciteBstWouldAddEndPuncttrue
\mciteSetBstMidEndSepPunct{\mcitedefaultmidpunct}
{\mcitedefaultendpunct}{\mcitedefaultseppunct}\relax
\EndOfBibitem
\bibitem[Raza \latin{et~al.}(2013)Raza, Stenger, Kadkhodazadeh, Fischer,
  Kostesha, Jauho, Burrows, Wubs, and Mortensen]{Raza:2013}
Raza,~S.; Stenger,~N.; Kadkhodazadeh,~S.; Fischer,~S.~V.; Kostesha,~N.;
  Jauho,~A.-P.; Burrows,~A.; Wubs,~M.; Mortensen,~N.~A. Blueshift of the
  surface plasmon resonance in silver nanoparticles studied with
  \uppercase{EELS}. \emph{Nanophotonics} \textbf{2013}, \emph{2},
  131--138\relax
\mciteBstWouldAddEndPuncttrue
\mciteSetBstMidEndSepPunct{\mcitedefaultmidpunct}
{\mcitedefaultendpunct}{\mcitedefaultseppunct}\relax
\EndOfBibitem
\bibitem[Raza \latin{et~al.}(2015)Raza, Kadkhodazadeh, Christensen, {Di Vece},
  Wubs, Mortensen, and Stenger]{Raza:2015}
Raza,~S.; Kadkhodazadeh,~S.; Christensen,~T.; {Di Vece},~M.; Wubs,~M.;
  Mortensen,~N.~A.; Stenger,~N. Multipole plasmons and their disappearance in
  few-nanometer silver nanoparticles. \emph{Nat. Commun.} \textbf{2015},
  \emph{6}, 8788\relax
\mciteBstWouldAddEndPuncttrue
\mciteSetBstMidEndSepPunct{\mcitedefaultmidpunct}
{\mcitedefaultendpunct}{\mcitedefaultseppunct}\relax
\EndOfBibitem
\bibitem[Mueller \latin{et~al.}(1990)Mueller, Miller, and Chiang]{MMC90}
Mueller,~M.~A.; Miller,~T.; Chiang,~T.-C. Determination of the bulk band
  structure of Ag in Ag/Cu (111) quantum-well systems. \emph{Phys.\ Rev.\ B}
  \textbf{1990}, \emph{41}, 5214\relax
\mciteBstWouldAddEndPuncttrue
\mciteSetBstMidEndSepPunct{\mcitedefaultmidpunct}
{\mcitedefaultendpunct}{\mcitedefaultseppunct}\relax
\EndOfBibitem
\bibitem[Matsuda \latin{et~al.}(2004)Matsuda, Tanikawa, Hasegawa, Yeom, Tono,
  and Ohta]{MTH04}
Matsuda,~I.; Tanikawa,~T.; Hasegawa,~S.; Yeom,~H.~W.; Tono,~K.; Ohta,~T.
  Quantum-Well States in Ultra-Thin Metal Films on Semiconductor Surfaces.
  \emph{e-J. Surf. Sci. Nanotechnol.} \textbf{2004}, \emph{2}, 169--177\relax
\mciteBstWouldAddEndPuncttrue
\mciteSetBstMidEndSepPunct{\mcitedefaultmidpunct}
{\mcitedefaultendpunct}{\mcitedefaultseppunct}\relax
\EndOfBibitem
\bibitem[Becker \latin{et~al.}(2006)Becker, Crampin, and Berndt]{BCB06}
Becker,~M.; Crampin,~S.; Berndt,~R. Theoretical analysis of STM-derived
  lifetimes of excitations in the Shockley surface-state band of Ag (111).
  \emph{Phys.\ Rev.\ B} \textbf{2006}, \emph{73}, 081402\relax
\mciteBstWouldAddEndPuncttrue
\mciteSetBstMidEndSepPunct{\mcitedefaultmidpunct}
{\mcitedefaultendpunct}{\mcitedefaultseppunct}\relax
\EndOfBibitem
\bibitem[Garc{\'i}a-Lekue \latin{et~al.}(2003)Garc{\'i}a-Lekue, Pitarke,
  Chulkov, Liebsch, and Echenique]{GPC03}
Garc{\'i}a-Lekue,~A.; Pitarke,~J.~M.; Chulkov,~E.~V.; Liebsch,~A.;
  Echenique,~P.~M. Role of occupied $d$ bands in the dynamics of excited
  electrons and holes in Ag. \emph{Phys.\ Rev.\ B} \textbf{2003}, \emph{68},
  045103\relax
\mciteBstWouldAddEndPuncttrue
\mciteSetBstMidEndSepPunct{\mcitedefaultmidpunct}
{\mcitedefaultendpunct}{\mcitedefaultseppunct}\relax
\EndOfBibitem
\bibitem[Kevan and Gaylord(1987)Kevan, and Gaylord]{KG1987}
Kevan,~S.~D.; Gaylord,~R.~H. High-resolution photoemission study of the
  electronic structure of the noble-metal (111) surfaces. \emph{Phys. Rev. B}
  \textbf{1987}, \emph{36}, 5809\relax
\mciteBstWouldAddEndPuncttrue
\mciteSetBstMidEndSepPunct{\mcitedefaultmidpunct}
{\mcitedefaultendpunct}{\mcitedefaultseppunct}\relax
\EndOfBibitem
\bibitem[Paniago \latin{et~al.}(1995)Paniago, Matzdorf, Meister, and
  Goldmann]{PMM95}
Paniago,~R.; Matzdorf,~R.; Meister,~G.; Goldmann,~A. Temperature dependence of
  Shockley-type surface energy bands on Cu (111), Ag (111) and Au (111).
  \emph{Surf.\ Sci.} \textbf{1995}, \emph{336}, 113--122\relax
\mciteBstWouldAddEndPuncttrue
\mciteSetBstMidEndSepPunct{\mcitedefaultmidpunct}
{\mcitedefaultendpunct}{\mcitedefaultseppunct}\relax
\EndOfBibitem
\bibitem[Michaelson(1977)]{M1977}
Michaelson,~H.~B. The work function of the elements and its periodicity.
  \emph{J.\ Appl.\ Phys.} \textbf{1977}, \emph{48}, 4729--4733\relax
\mciteBstWouldAddEndPuncttrue
\mciteSetBstMidEndSepPunct{\mcitedefaultmidpunct}
{\mcitedefaultendpunct}{\mcitedefaultseppunct}\relax
\EndOfBibitem
\end{mcitethebibliography}
\end{document}